\documentclass{article}
\usepackage[utf8]{inputenc}
\usepackage[T1]{fontenc}
\usepackage{arxiv} 
\usepackage{lmodern}
\usepackage{microtype}

\usepackage{amsmath,amssymb,amsfonts}
\usepackage{bm}
\usepackage{siunitx}

\usepackage{booktabs}
\usepackage{longtable}
\usepackage{footnote}
\makesavenoteenv{longtable}

\usepackage{graphicx}
\usepackage{caption}
\usepackage{float}

\usepackage{xcolor}

\usepackage{natbib}
\usepackage{doi}
\usepackage{xurl}
\usepackage{etoolbox}
\usepackage{hyperref}

\makeatletter
\patchcmd\longtable{\par}{\if@noskipsec\mbox{}\fi\par}{}{}
\makeatother
\makeatletter
\def\maxwidth{\ifdim\Gin@nat@width>\linewidth\linewidth\else\Gin@nat@width\fi}
\def\maxheight{\ifdim\Gin@nat@height>\textheight\textheight\else\Gin@nat@height\fi}
\makeatother
\setkeys{Gin}{width=\maxwidth,height=\maxheight,keepaspectratio}

\title{The Energy Footprint of LLM-Based Environmental Analysis: LLMs and Domain Products}

\author{
\parbox[t]{0.30\textwidth}{\centering
\textbf{Alicia Bao}\thanks{Authors contributed equally to this work and are listed alphabetically.} \\
Department of Computer Science\\
University of North Carolina at Chapel Hill\\
Chapel Hill, NC 27516\\
\texttt{alicia.bao@unc.edu}
}
\And
\parbox[t]{0.30\textwidth}{\centering
\href{https://orcid.org/0009-0005-3397-6741}{\includegraphics[scale=0.06]{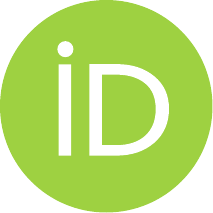}\hspace{1mm}\textbf{Jiamian He}} \\
Arboretica\\
Rotterdam, The Netherlands\\
\texttt{jiamian@arboretica.com}
}
\And
\parbox[t]{0.30\textwidth}{\centering
\href{https://orcid.org/0000-0003-4913-9479}{\includegraphics[scale=0.06]{orcid.pdf}\hspace{1mm}\textbf{Angel Hsu}} \\
Department of Public Policy\\
University of North Carolina at Chapel Hill\\
Chapel Hill, NC 27516\\
\texttt{angel.hsu@unc.edu}
}
\AND
\parbox[t]{0.30\textwidth}{\centering
\href{https://orcid.org/0000-0001-8429-0954}{\includegraphics[scale=0.06]{orcid.pdf}\hspace{1mm}\textbf{Diego Manya}} \\
Data-Driven EnviroLab\\
UNC Institute for Environment\\
Chapel Hill, NC 27516\\
\texttt{diego.manya@unc.edu}
}
\And
\parbox[t]{0.30\textwidth}{\centering
\href{https://orcid.org/0009-0003-7588-1404}{\includegraphics[scale=0.06]{orcid.pdf}\hspace{1mm}\textbf{Ji (James) Zhang}} \\
Arboretica\\
Rotterdam, The Netherlands\\
\texttt{james@arboretica.com}
}
}

\renewcommand{\shorttitle}{\textit Energy consumption of LLMs}

\hypersetup{
pdftitle={Energy footprint of LLM analysus},
pdfsubject={q-bio.NC, q-bio.QM},
pdfauthor={Alicia Bao, Jiamian He, Angel Hsu, Diego Manya, Ji (James) Zhang},
pdfkeywords={LLMs, energy consumption, domain products},
}

\begin{document}
\maketitle

\begin{abstract}
As large language models (LLMs) are increasingly used in domain-specific applications, including climate change and environmental research, understanding their energy footprint has become an important concern. The growing adoption of retrieval-augmented (RAG) systems for climate-domain specific analysis raises a key question: how does the energy consumption of domain-specific RAG workflows compare with that of direct generic LLM usage? Prior research has focused on standalone model calls or coarse token-based estimates, while leaving the energy implications of deployed application workflows insufficiently understood. In this paper, we assess the inference-time energy consumption of two LLM-based climate analysis chatbots (ChatNetZero and ChatNDC) compared to the generic GPT-4o-mini model. We estimate energy use under actual user queries by decomposing each workflow into retrieval, generation, and hallucination-checking components. We also test across different times of day and geographic access locations. Our results show that the energy consumption of domain-specific RAG systems depends strongly on their design. More agentic pipelines substantially increase inference-time energy use, particularly when used for additional accuracy or verification checks, although they may not yield proportional gains in response quality. While more research is needed to further test these initial findings more robustly across models, environments and prompting structures, this study provides a new understanding on how the design of domain-specific LLM products affects both the energy footprint and quality of output.
\end{abstract}


\newpage

\section{Introduction}
While the rapid proliferation of Generative Artificial Intelligence
(GenAI) and Large Language Model (LLM)-powered chatbots represents a
transformative shift in how information is generated and accessed, major
questions remain regarding the energy-intensity and environmental
footprint of utilizing these tools compared to other means of
information acquisition, such as traditional Internet search tools. One
approach that has been touted as a potential solution to very large,
generic LLM-based chatbots that have been trained with massive amounts
of data \cite{samborska_scaling_2025} is the development of Domain-Specific
Retrieval-Augmented generation (RAG) chatbots that represent a
``smaller'' (small-LLM) \cite{vrettos_accurate_nodate}  alternative.
Particularly in the environment, climate and sustainability domains,
domain-specific RAG chatbots have emerged, promising more accurate
information related to climate science and policy compared to generic
chatbots \cite{vaghefi_chatipcc_2023, hsu_evaluating_2024, shahbazi_using_2026}. While these tools claim higher accuracy and
``anti-hallucination'' filters, what is less transparent is whether the
proposed climate benefits of these tools may be paradoxically offset
from their energy consumption and environmental impacts.

The question of GenAI's energy and environmental footprints has risen in
scrutiny and public concern, given the increasing number of data centers
planned or in construction across the U.S. as a result of AI's growing
needs for training and running models \cite{meredith_thirsty_2023, copley_data_2025}. Although data centers are estimated to only account for 1\% of
total global electricity demand \cite{iea_energy_2025}, this percentage is
projected to grow to as much as 12\% in 2030 \cite{chen_how_2025}. Prior
research has demonstrated that while model training is technically much
more energy-intensive than inference \cite{kaack_aligning_2022}, the inference
phase in aggregate is responsible for the vast majority of AI's energy
consumption today given the scale of AI usage \cite{odonnell_we_2025}. In
light of rising public concern, hyperscaler AI companies like Microsoft
and Google have released estimates of the water, energy and
environmental impacts of AI usage \cite{elsworth_measuring_2025}, although critics
have pointed out the lack of transparency and potential inaccuracies in
these estimates \cite{li_making_2025}.

Here, we present a methodological refinement of a prior measurement
framework \cite{jegham_how_2025}, enabling an empirical end-to-end carbon
emission assessment of real-world, domain-specific LLM-RAG products used
for environmental analysis. We introduce a query-level methodology to
decompose and quantify the energy consumption of RAG sub-components,
rigorously benchmarking the relative costs of the Retrieval and the
Generation stages. Ultimately, our findings yield the first quantitative
framework for evaluating this complex environmental trade-off (energy
consumption vs quality), leading to actionable, data-driven
architectural guidelines for designing demonstrably more sustainable
generative AI systems.

\section{Background}
\subsection{ChatNetZero and ChatNDC - Climate-domain specific
chatbots}
\href{http://chatnetzero.ai}{ChatNetZero.ai} (CNZ) is a climate-specific
chatbot designed to reduce the well-known limitations of generic LLMs,
especially hallucination, by pairing a RAG pipeline with purpose-built
validation and referencing modules \cite{hsu_evaluating_2024}. In the model,
heterogeneous data sources (e.g., PDFs and spreadsheets) are converted
to plain text, segmented into paragraph-preserving chunks, and embedded
alongside user queries to enable semantic retrieval. Each chunk is
tagged with document and page metadata to enable semantic retrieval to
preserve provenance. CNZ then applies query processing that
distinguishes actor-specific questions (e.g., named governments or
companies) from generic questions, prioritizing Net Zero Tracker \cite{net_zero_tracker_net_2025} entities when detecting actor-specific contexts.
CNZ then retrieves a bounded set of highly relevant chunks per
actor while using top-ranked report chunks for non-actor queries.
Retrieved evidence is combined with the user prompt and passed to an LLM
under tightly constrained prompting (e.g., temperature set to 0 and
explicit instructions to answer only from provided text and acknowledge
uncertainty). To further mitigate hallucination, an anti-hallucination
module embeds the draft output sentence-by-sentence and verifies each
claim against retrieved source chunks, excluding sentences that cannot
be traced back to the underlying corpus. A reference module then appends
per-sentence citations, including document name, page number, and
matched location, linking users directly to original source pages for
manual verification. Finally, CNZ extends beyond narrative Q\&A
by transforming structured Net Zero Tracker variables into
natural-language statements, enabling quantitative and counting-style
queries to be handled through the same retrieval-and-grounding workflow.

\begin{figure}[t]
\centering
\includegraphics[height=3.5in]{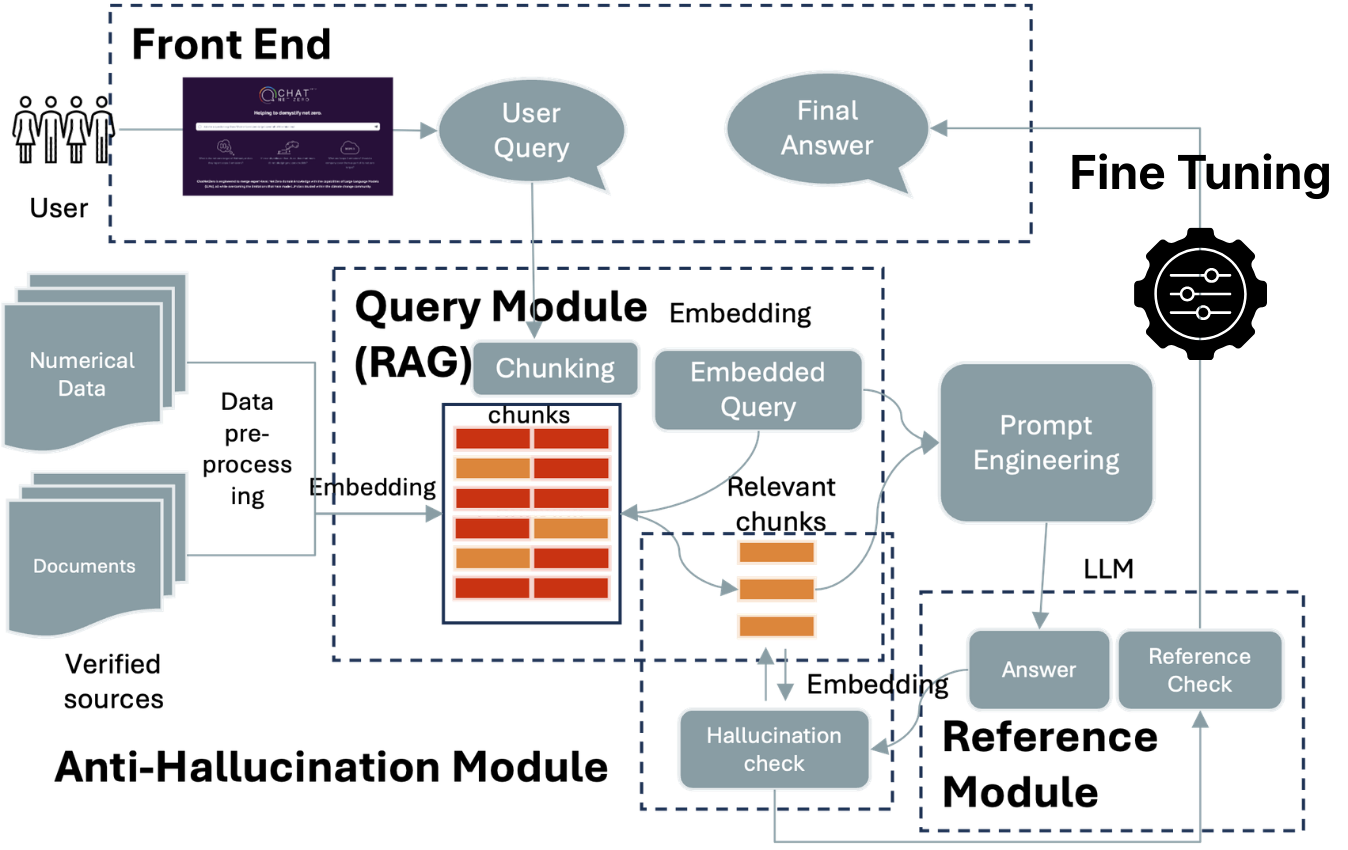}
\caption{Design of ChatNetZero \cite{hsu_evaluating_2024}.}
\label{fig:ChatNetZero}
\end{figure}

ChatNDC (ChatNDC.org, beta) is a climate policy AI chatbot launched in
January 2025 by Data-Driven EnviroLab and Arboretica to streamline
credible analysis of countries' Nationally Determined Contributions
(NDCs) under the Paris Agreement. The system is grounded in a curated
corpus of verified climate documents spanning officially submitted
national policy texts reported to the The United Nations Framework
Convention on Climate Change (UNFCCC), country-specific scientific and
policy assessments, and global syntheses on climate science, finance,
mitigation, and adaptation (including the latest IPCC assessments),
enabling users to interrogate emissions targets, timelines, sectoral
priorities, and changes across successive NDC iterations. To minimize
hallucinations, ChatNDC pairs generation with an ``anti-hallucination''
logic that cross-verifies outputs against trusted sources in real time
and returns granular, auditable references (both embedded in the
narrative response and linked back to precise locations in the
underlying documents) for users to directly validate claims. Beyond
question and answering capabilities, the platform supports
document-level exploration via a data catalog, allowing users to query
one or multiple source documents and receive evidence-forward responses
that surface and highlight the relevant text passages, effectively
turning complex climate reports into interactive analytical objects. In
practice, this grounding-and-referencing approach improves reliability
over generic chatbots. For example, on questions such as whether Egypt
has declared a net-zero target, where unconstrained LLMs may produce
false positives, ChatNDC was designed to provide a defensible answer
with traceable citations to Egypt's submitted materials.

\subsection{LLM energy consumption components: training,
inference}
For the purposes of evaluating LLM-based chatbots, operational energy
consumption is comprised primarily of two main phases: training and
inference \cite{luccioni_power_2024}. Training includes pre-training and
any subsequent fine-tuning and is typically concentrated in short but
highly compute-intensive runs \cite{chen_how_2025}. The energy demand
during this phase is determined by a range of factors, including model
parameter count, the size of the training dataset, processor, hardware
Thermal Design Power (TDP), and the Power Usage Effectiveness (PUE) of
the data center's cooling infrastructure \cite{patterson_carbon_2021}. For instance, it is
estimated that the training of GPT-4 consumed around 50 GWh of
electricity \cite{ariyanti_sri_trade-off_nodate}. Similarly, while Google's
Gemini series employs more energy-efficient architectures and
specialized Tensor Processing Units (TPUs), the training cost for LLMs
remains at a comparable volume, equivalent to the annual energy
consumption of thousands of average households \cite{cottier_rising_2025}. Although these figures
represent an initial surge in electricity demand, the total energy
consumption becomes fixed once the model is deployed.

In contrast, the inference phase begins after deployment and covers the
electricity required to process prompts and generate responses in real
time. This distinction is analytically important because the two phases
differ not only in temporal profile but also in their energy
consumption, since training usually happens over a short, discrete
period of time, whereas the inference (e.g., user phase) is continuous
and scales directly with usage volume. Evidence indicates that for a
model of GPT-3's scale, once daily active users reach the tens of
millions, it takes a mere 43 days of continuous operation for cumulative
inference energy to equal the initial training energy \cite{guardia_assessing_2024}. For popular models, this
\lq energy parity point\rq \space is reached even more rapidly, as the computational
complexity of inference increases non-linearly with the length of the
input context \cite{luccioni_power_2024}.

Early work on the environmental cost of LLMs and AI emphasized the
substantial resource demands of the training phase, while more recent
work has argued that inference deserves equal or greater attention once
models are deployed across user-facing applications like chatbots
\cite{lacoste_quantifying_2019, kaack_aligning_2022, luccioni_power_2024}.

\subsection{Previous research on measuring LLM inference
methods}
To quantify carbon emissions during the LLM inference phase, current
research primarily adopts three methods: local hardware-based
monitoring, architecture-driven parameterized estimation, and end-to-end
API-based estimation.

\textbf{Hardware-based Monitoring (Local Deployment)}: The most
prevalent approach involves direct energy measurement via hardware-level
sensors (GPU, RAM, CPU) or Cloud Service Provider (CSP) embedded
calculators. This method is favored for its high precision and
controllability within laboratory environments. Consequently, such
studies \cite{everman_evaluating_2023, dauner_energy_2025, nguyen_towards_2024} typically deploy open-source models for
measurement.

\textbf{Architecture-driven Parameterized Estimation}: Analytical tools,
such as LLMCarbon \cite{faiz_llmcarbon_2024} and $\mathrm{LLMCO_2}$ \cite{fu_llmco2_2025}, utilize parameterized models to estimate energy consumption and
carbon emissions. These techniques rely heavily on specific model
architecture details to compute Floating Point Operations (FLOPs).
Furthermore, to bridge the gap between theoretical FLOPs and actual
consumption, the R-ICE \cite{sikand_breaking_2025} research trains
regression models within the framework, creating benchmarks that more
accurately map inference carbon and energy metrics across diverse
hardware configurations.

\textbf{End-to-End API-based Estimation}: For commercial `black-box'
models, where architectural details and local deployment are limited,
end-to-end API estimation has emerged as an efficient evaluation path.
Related studies include the MELODI framework \cite{husom_price_2026},
and the subsequent work by Jegham et al. \cite{jegham_how_2025} both
of which systematically evaluate the environmental impacts of LLM
inference. Jegham et al.~\cite{jegham_how_2025} propose a framework that estimates energy usage
based on three core components: inference latency, estimated model/GPU
scale, and PUE. This method has the
generalizability that it represents a new trend, a viable approach for
conducting end-to-end assessments of commercial black-box models.

\subsection{Limitations of previous
research}
Previous approaches to estimating LLM energy consumption remain limited
in ways that reduce their usefulness for evaluating deployed chatbot
products. Local measurement methods generally require direct access to
the model runtime and hardware, which makes them best suited to locally
hosted, open-weight systems rather than proprietary API-based models. As
a result, much of the existing empirical literature has focused on
smaller (e.g., less than 40 billion parameters) or open-source models,
while widely used closed models such as GPT-4o remain difficult to
assess directly because their architectures, deployment configurations,
and hardware allocations are not publicly disclosed. In these settings,
FLOPs-based analytical methods can be informative, but they depend on
assumptions about model structure and infrastructure that are often
unverifiable in commercial deployments. Prior reviews similar note that
local prompt-level methods have largely concentrated on open-source or
small-scale settings and do not fully capture the infrastructure
complexity of production inference \cite{jegham_how_2025}.

Recent API-based estimation methods help address part of this gap. In
particular, Jegham et al.~\cite{jegham_how_2025} propose an infrastructure-aware
framework that combines public API latency and throughout data with
published hardware specifications and environmental multipliers to
estimate prompt-level inference impacts for proprietary as well as open
models. Their study represents an important advance because it moves
beyond purely local measurement and makes closed commercial systems more
tractable for sustainability analysis. This approach is also designed
primarily to estimate the environmental footprint of individual model
calls across standardized query-size categories, rather than the full
execution logic of application-layer systems. Their ``How Hungry is AI''
dashboard relies on daily-scraped API performance data from Artificial
Analysis, which allows for cross-model
benchmarking, but is less suited to reconstructing the exact sequence of
operations performed by a specialized RAG product for a particular user
query \cite{jegham_how_2025}.

This distinction matters because in practice, real-world RAG systems,
especially those used with agentic workflows, could trigger multiple LLM
API calls within a single user request, with the number and structure of
calls determined by the system design rather than token length alone.
Agentic or multi-stage pipelines therefore introduce energy costs that
cannot be inferred reliably from prompt size categories or from a single
end-to-end API call. Additionally, different pipeline components rely on
different computational resources: retrieval may be primarily CPU-bound,
whereas generation and LLM-based hallucination checks depend more
heavily on GPU-backed inference. For this reason, accurately estimating
inference-time energy consumption for deploying RAG system requires a
pipeline decomposition that reflects actual execution patterns under
real queries.

Another limitation of prior work is that energy use is often quantified
in isolation from output quality or accuracy. Existing studies have
substantially improved the understanding of LLM inference's resource
demands, but they generally do not examine whether additional
computational cost translates into better answers, stronger retrieval
grounding, or reduced hallucination in applied settings. These gaps
leave an important practical question unresolved: whether more
energy-intensive workflows deliver meaningful gains in response quality,
especially when systems add extra inference steps for routing,
validation or reference checking.

This present study addresses these gaps by shifting the unit of analysis
from the standalone model call to the deployed application workflow.
Rather than estimating energy use from coarse token-length bands alone,
we empirically quantify the energy consumption of domain-specific LLM
and RAG products under actual user queries in a specialized
environmental context. We further decompose each RAG workflow into its
constituent inference-time steps, including query classification,
retrieval, response generation, and hallucination checking, and estimate
the energy associated with each component separately. This step-level
approach makes it possible to identify where energy is actually consumed
within LLM-based chatbot systems, including components that would be
obscured in model-level averages. In addition, to improve the robustness
of our estimates, we test energy consumption across different times of
day and geographic locations by varying IP addresses, thereby capturing
temporal and spatial variation in API response behavior and producing
more reliable average energy estimates across systems and user contexts.
Finally, by evaluating both energy use and response quality, our
framework makes visible the trade-offs between computational cost and
performance that are central to the real-world design of LLM-based
applications. While foundational studies, particularly Jegham et al.~\cite{jegham_how_2025}, provide important advances in benchmarking the electricity
consumption of LLM APIs through an infrastructure-aware framework that
integrates public API performance data, region-specific environmental
multipliers, and inferred hardware configurations, our study extends
this literature by moving from model-level, token-range-based estimation
to the direct empirical evaluation of deployed LLM and RAG applications
under real queries.

\section{Methods}
\subsection{\texorpdfstring{Dataset}{Dataset }}
To evaluate energy consumption and response quality under realistic
usage conditions, we constructed a dataset of 102 domain-specific test
questions reflecting the kinds of queries users are likely to ask in
deployed climate-LLM applications. The dataset was manually curated to
cover both direct factual retrieval and higher-order analytical tasks
within the climate policy domain. Substantively, the questions span
corporate climate commitments, net-zero targets, offset conditions, and
national climate policy topics, including country's Paris Agreement
pledges or NDCs, for which the
\href{http://chatndc.org}{ChatNDC.org} platform was designed to
analyze. We classified the questions following Bloom\textquotesingle s
Taxonomy \cite{ormell_blooms_1974}: Knowledge, Comprehension, Application,
Analysis, Evaluation and Creation with 48 questions of the knowledge
class and approximately 10 of each of the other classes. We particularly
focused on the Knowledge class due to their particular alignment with
the use-case of our custom models (CNZ and ChatNDC) as well as
for the considerable breadth of questions that could be included in this
category such as questions that just retrieve data as well as those that
summarize it. Figure \ref{fig:med_energy} shows the median and variability of the token
size for each answer by each model, highlighting the significant
variability in output token from unconstrained models such as GPT
4o-Mini, and the very limited variance in answer length from constrained
model such as GPT 4o-Mini (200) and CNZ. A subset of these
questions, between 19 to 25 for each model, was selected for manual
evaluation on their factual accuracy and embellishment prevalence.

\begin{figure}[t]
\centering
\includegraphics[width=3.02604in,height=3.02604in]{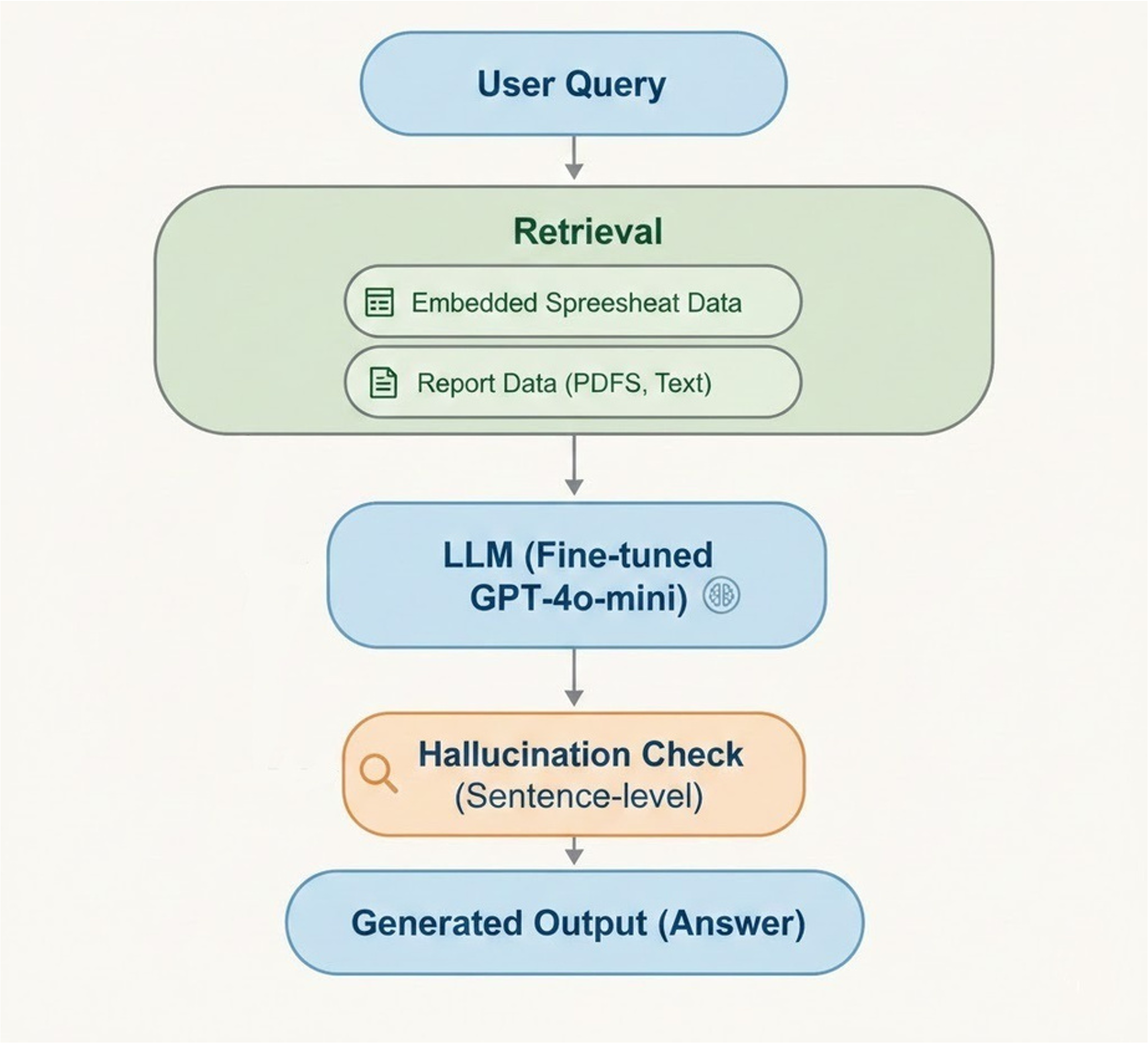}
\includegraphics[width=3.02604in,height=3.02604in,clip,trim=2pt 2pt 2pt 2pt]{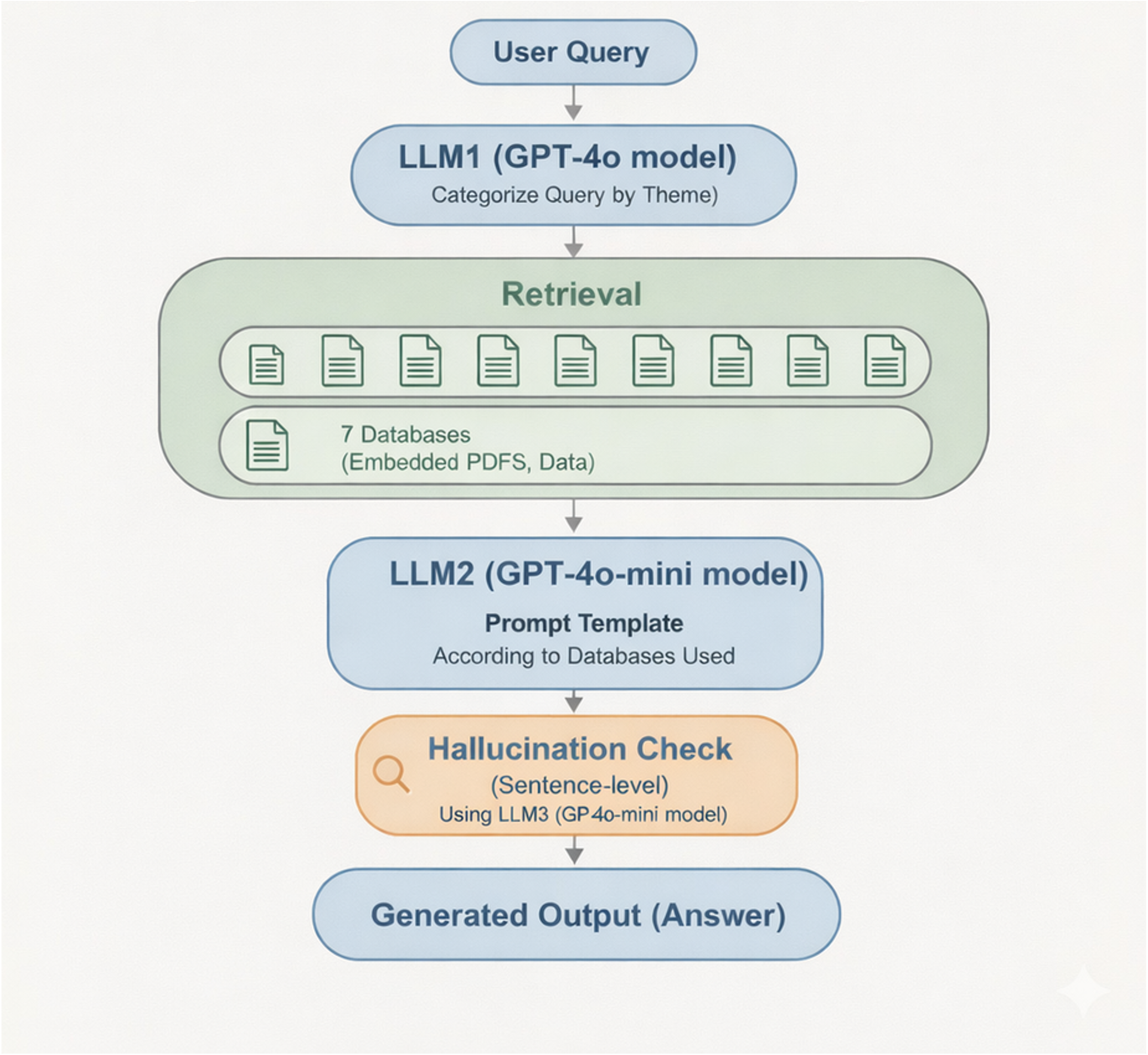}
\caption{Architectures of (a) ChatNetZero and (b) ChatNDC.}
\label{fig:CNZ_vs_NDC}
\end{figure}

\renewcommand{\arraystretch}{1.15} 
\setlength{\LTpre}{6pt}           
\setlength{\LTpost}{6pt}          

\begin{longtable}{@{}p{0.15\linewidth}p{0.81\linewidth}@{}}
\caption{Representative questions from the evaluation dataset.}\label{tab:rep_questions}\\
\toprule
\textbf{Question type} & \textbf{Example questions} \\
\midrule
\endfirsthead

\toprule
\textbf{Question type} & \textbf{Example questions} \\
\midrule
\endhead

Knowledge &
``Has Marathon Petroleum made any climate pledges?''; ``Does Foxconn have any climate strategy?''; ``What is the total emissions reduction target of Malaysia?''; ``Does Taiwan have a carbon trading scheme or policy under its latest NDC?'' \\[4pt]

Comprehension &
``Interpret how ESG regulations influence business strategy and investor decisions''; ``Explain the relationship between local air pollution and public health outcomes.'' \\[4pt]

Application &
``What are Scope 3 emissions and what should be included in a net-zero target?''; ``If an entity doesn't control out-of-boundary emissions, how can it credibly set a net-zero target?'' \\[4pt]

Analysis &
``Are Costco's or Walmart's net zero target 1.5 degree aligned?''; ``How does Walmart's climate goals compare with other large retail stores?''; ``How do England and Scotland NDCs compare to the UK's?'' \\[4pt]

Evaluation &
``Is Apple's net zero target credible?''; ``How is supermarket chain Jeronimo Martins doing on their commitment to reduce food waste by 50\% by 2025?'' \\[4pt]

Creation &
``Propose an ESG audit model that integrates social, environmental, and ethical performance indicators into one scorecard.''; ``Develop a set of innovative KPIs to measure corporate progress toward net-zero goals beyond traditional carbon metrics.'' \\

\bottomrule
\end{longtable}

\subsection{Energy Consumption
Estimation}
\subsubsection*{\texorpdfstring{ 3.2.1 ChatNetZero \& ChatNDC structure
}{ 3.2.1 ChatNetZero \& ChatNDC structure }}\addcontentsline{toc}{subsubsection}{ 3.2.1 ChatNetZero \& ChatNDC
structure }

Unlike a direct API call for LLMs, different domain RAG systems have
different structures. CNZ uses a simple RAG as shown in Figure \ref{fig:ChatNetZero}. It
mainly contains one retrieval, one LLM call with a self-fine-tuned GPT
4o-mini model, and uses the pairwise cosine similarity method to check
the LLM's raw response relevance by sentence with the retrieval
documents. If the cosine score is smaller than a threshold, the sentence
in response will be deleted from the final output. We use cosine
similarity to determine how semantically similar two texts are by
measuring the cosine of the angle between their vector representations.
In our pipeline, a low cosine score indicates that a generated sentence
is semantically inconsistent (potentially indicating a hallucination)
compared to existing content, triggering its removal to ensure a concise
output.

ChatNDC is a more complex RAG, with two more LLM inference calls than
CNZ. As shown in Figure \ref{fig:CNZ_vs_NDC}, the system first uses an LLM
inference (GPT-4o) to categorize the user query by theme, which
determines the relevant databases for retrieval. The retrieved content
is then passed to the second LLM inference call (GPT 4o-Mini) with a
database-aware prompt template to generate the response. Finally, a
third LLM inference call (GPT 4o-Mini) performs a sentence-level
hallucination check to validate the generated content before outputting
the final answer. Compared to the CNZ cosine similarity method, which
relies on a static threshold value, ChatNDC employs an LLM-based
approach for hallucination checks. This choice is driven by the LLM's
flexible reasoning capability and its automated citation verification
(cross-referencing) ability. Unlike fixed mathematical metrics, the LLM
can intelligently assess whether a generated statement is accurately
supported by the retrieved references, ensuring both factual integrity
and logical consistency.

Thus, for the RAG systems, the total energy consumption ($E_{total}$) is
modeled by aggregating the energy costs across three main inference
parts of the RAG systems:

\begin{itemize}
\item
  Retrieval: retrieve the documents in the vector database by query
  (Spreadsheets or PDFs).
\item
  LLM Inference: API request to completion of the generated answer.
\item
  Hallucination Check: pairwise algorithm or LLM for the sentence-level
  consistency verification.
\end{itemize}

The total energy consumption is represented by: $E_{total}$ = $E_{retrieval}$
+ $E_{inference}$ + $E_{hallucination}$.

\begin{figure}[t]
\centering
\includegraphics[height=2.5in,clip,trim=6pt 6pt 6pt 6pt]{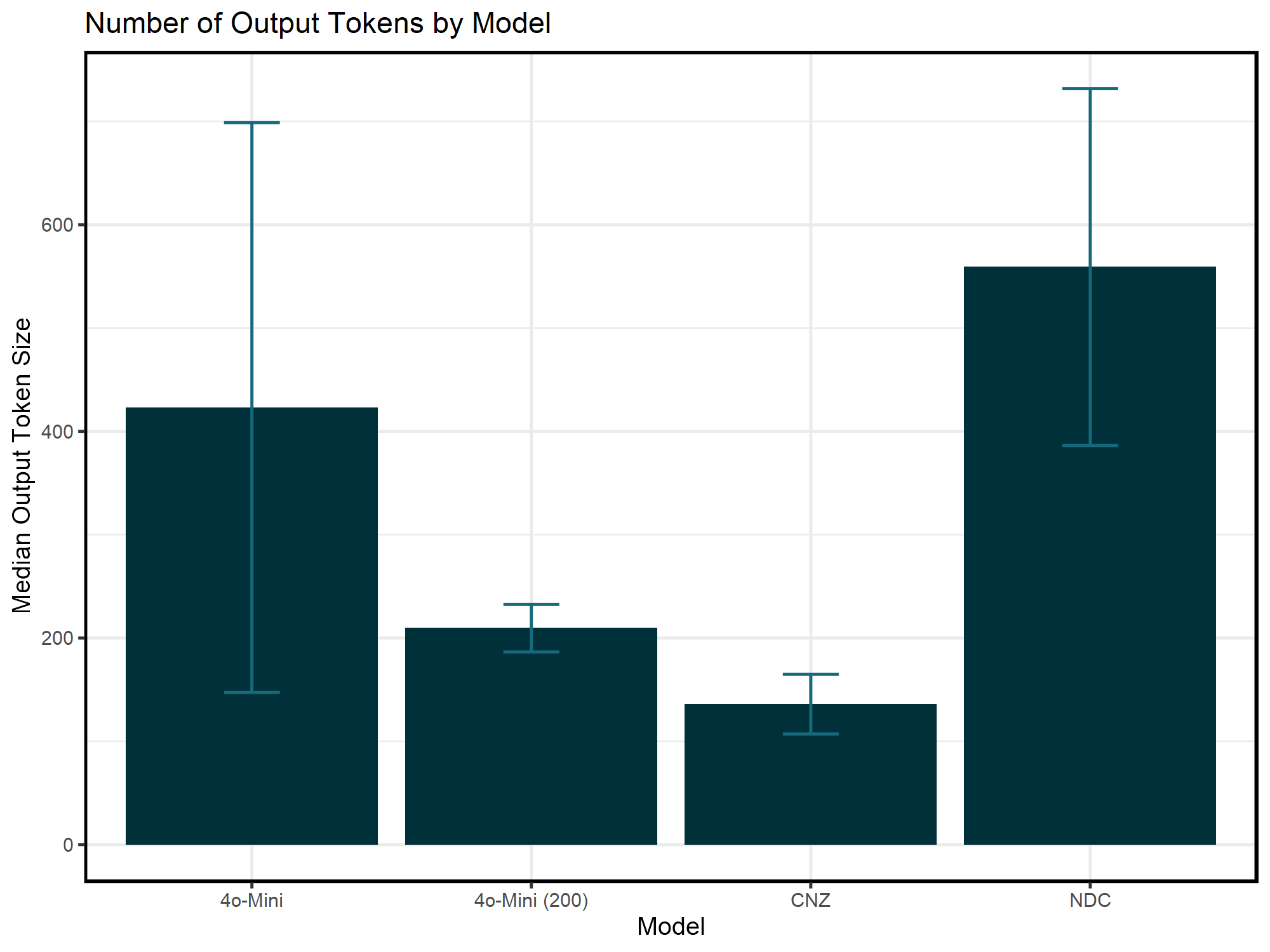}
\caption{Median answer token size by each model.}
\label{fig:token_output}
\setlength{\textfloatsep}{8pt}
\end{figure}

\subsubsection{Experiment Design}
Our objective is to quantify the inference-time and energy consumption
of different LLM-based chatbot pipelines by measuring execution time and
decomposing the computational steps within each workflow. To do so, we
conduct controlled experiments that isolate the major components of each
pipeline and evaluate their performance under variation in time of day.
This design allows us to generate a more representative estimate of
average energy consumption by accounting for fluctuations in latency
associated with network conditions and server load. Detailed
energy-estimation procedures are described in Sections 3.2.3--3.2.5.

All experiments were conducted in a Google Colab environment using
Python 3 and a single-core CPU runtime. All LLMs were accessed through
API calls with the temperature parameter fixed at 0. Setting the
`temperature' to 0 reduces stochastic variation by making generation
deterministic, thereby improving reproducibility and minimizing noise in
repeated measurements of latency, token use, and estimated energy
consumption.

Because both CNZ and ChatNDC use GPT 4o-Mini as their primary generation
model, we restrict comparison to four pipelines: CNZ, ChatNDC,
GPT 4o-Mini, and GPT 4o-Mini (200). The GPT 4o-Mini (200) condition
imposes a maximum output length of 200 words in order to provide a
controlled baseline for comparison with CNZ, whose responses are
inherently constrained to this approximate length. This shared
base-model design provides a consistent benchmark and allows us to
isolate differences attributable to workflow structure rather than
differences in model family, parameter scale, or provider
infrastructure. Although absolute energy consumption can vary across
other model architectures and hosting environments, holding the
underlying model constant improves comparability across our experimental
conditions.

The experiment consists of two complementary parts, both based on the
same evaluation dataset. The first part focuses on energy estimation
robustness. To account for variation in network latency and API server
load, each pipeline was executed repeatedly across four runs: one run at
a randomly selected time and three runs during fixed time windows in the
morning (08:00--10:30), afternoon (14:00--16:30), and evening
(20:00--22:30). All runs originated from the Netherlands. For each run,
we recorded complete execution time and token usage and converted these
measurements into estimated energy consumption. Query and response
tokens were counted using the tiktoken library
\cite{noauthor_tiktoken_2026} configured for GPT 4o-Mini.

Finally, we combine estimated cumulative energy consumption with
response quality evaluation to examine the trade-offs between
computational cost and performance across systems. By jointly examining
energy use, token consumption and answer quality, we assess the relative
efficiency of each pipeline and identify whether additional workflow and
RAG design complexity yield meaningful improvements in output quality.

\subsubsection{LLM consumption estimation
method}
Building upon Jegham et al.'s methodology ~\cite{jegham_how_2025}, we refine their
approach to measure LLM energy consumption. As shown in Formula 1, $E_{query}$ represents the total energy
consumption. The total inference time, $T_{total}$ (the sum of terms in the first
bracket), corresponds to the elapsed time from calling the API until the
final full answer is obtained. The total workload, $W_{total}$ (the sum of terms in the second
bracket), is calculated according to Formula 2. The PUE is the 1.12 as
mentioned in the Microsoft datacenter report \cite{microsoft_measuring_nodate}
where OpenAI holds their LLM models.

\begin{equation}
E_{\text{query}}\,(\text{kWh}) =
\left(\frac{\frac{\text{Output Length}}{\text{TPS}} + \text{Latency}}{3600}\right)
\left(P_{\text{GPU}} \times U_{\text{GPU}} + P_{\text{non-GPU}} \times U_{\text{non-GPU}}\right)
\cdot \text{PUE}
\end{equation}

\begin{equation}
U_{\text{GPU}} = \frac{G \times D_{\text{GPU}}}{N \times B},
\qquad
U_{\text{non-GPU}} = \frac{G \times D_{\text{non-GPU}}}{N \times B}.
\end{equation}

Here, $G$ denotes the number of GPUs assigned per model, $N$ the number of
GPUs per node, and $B$ is the batch size. $D_{\mathrm{GPU}}$ represents the assigned
power draw of GPUs, expressed as a fraction of their maximum rated
power, while $D_{\mathrm{non-GPU}}$ corresponds to a conservatively estimated fixed
utilization fraction for non-GPU components (e.g., CPU, memory, storage,
cooling). The LLM parameters, as reported in the reference paper, are
summarized below:

\begin{table}[ht]
\centering
\renewcommand{\arraystretch}{1.1}
\begin{tabular}{@{}llllllll@{}}
\toprule
\textbf{Model} & \textbf{Size} & \textbf{GPU} & \bm{$G$} & \bm{$D_{\mathrm{gpu}}$} & \bm{$D_{\mathrm{nongpu}}$} & \bm{$N$} & \bm{$B$} \\
\midrule
GPT-4o      & Large  & H100 & 8 & 0.6 & 0.5 & 8 & 8 \\
GPT 4o-Mini & Medium & A100 & 4 & 1.2 & 0.5 & 8 & 8 \\
\bottomrule
\end{tabular}
\end{table}

Therefore we estimate $W \cdot \mathrm{PUE}$ according to the Equations 1 and 2, and the
$T_{total}$ is obtained from the previously outlined and then we estimate the
parameters of the formula, presented in the Table below:

\begin{table}[ht]
\centering
\renewcommand{\arraystretch}{1.1}
\begin{tabular}{@{}llllllll@{}}
\toprule
\textbf{Model} & \bm{$U_{\mathrm{GPUtotal}}$} & \bm{$U_{\mathrm{GPU}}$} & \bm{$P_{\mathrm{GPU}}$} & \bm{$P_{\mathrm{non-GPU}}$} & \bm{$W_{\mathrm{total}}$} & \textbf{PUE} & \bm{$W \cdot \mathrm{PUE}$} \\
\midrule
GPT-4o      & 0.42 & 0.2875   & 5.6 & 4.6 & 0.7075   & 1.12 & 0.7924 \\
GPT 4o-Mini & 0.24 & 0.103125 & 3.2 & 3.3 & 0.343125 & 1.12 & 0.3843 \\
\bottomrule
\end{tabular}
\end{table}

\begin{equation}
E_{\mathrm{GPT\text{-}4o}}\,\mathrm{(kWh)}
=
T_{\mathrm{llm}}\,\mathrm{(h)}
\cdot 0.7075\,\mathrm{(kW)}
\cdot 1.12
=
T_{\mathrm{llm}}\,\mathrm{(h)}
\cdot 0.7924\,\mathrm{(kW)}
\end{equation}

\begin{equation}
E_{\mathrm{GPT\text{ }4o\text{ }mini}}\,\mathrm{(kWh)}
=
T_{\mathrm{llm}}\,\mathrm{(h)}
\cdot 0.343125\,\mathrm{(kW)}
\cdot 1.12
=
T_{\mathrm{llm}}\,\mathrm{(h)}
\cdot 0.3843\,\mathrm{(kW)}
\end{equation}

\subsubsection{Retrieval consumption estimation
method}
In the retrieval part, since all processes are executed within a Google
Colab environment using a single CPU core (typically an
Intel(R)-Xeon(R)-CPU@2.20GHz), the retrieval energy consumption is
approximated as the energy draw of a single CPU core. The official
specifications for this Intel(R)-Xeon(R)-CPU series \cite{intel_intel_nodate}
list several variants with TDP values ranging
from 150W to 330W, with a number of cores ranging from 12 to 44, but the
Colab platform does not specify the exact sub-model utilized during the
session. For cross-verification, we adopt a power consumption standard
of about 8.5W per core based on LLMCarbon data to serve as our primary
calculation benchmark. Furthermore, to accurately reflect the total
energy footprint, including infrastructure overhead, we incorporate a
PUE value of 1.09, as disclosed in the
Google Data Center Efficiency Report \cite{google_power_nodate}. The final energy consumption
for this phase is thus determined by the product of the measured
execution time, the power per core, and the PUE factor as follows:

\begin{equation}
E_{\mathrm{retrieval}}
=
T_{\mathrm{retrieval}} \cdot P_{\mathrm{CPU}} \cdot \mathrm{PUE}
\end{equation}

\begin{equation}
E_{\mathrm{retrieval}} \, (\mathrm{kWh})
=
T_{\mathrm{retrieval}} \, (h) \cdot 0.0085 \, (\mathrm{kW}) \cdot 1.09
=
T_{\mathrm{retrieval}} \, (h) \cdot 0.009265 \, (\mathrm{kW})
\end{equation}

\subsubsection{Hallucination Check consumption estimation
method}
As described above in section 3.2.1, CNZ
and NDC employ different hallucination-checking approaches, and these
differences are reflected in how their energy consumption is estimated.
In CNZ, hallucination checking is implemented through sentence-level
cosine similarity between each generated sentence and the retrieved
source documents. Because this process does not require an additional
LLM call, its energy use is attributed to the retrieval/post-processing
stage and accounted for within \(E_{\mathrm{retrieval}}\). In contrast,
ChatNDC performs hallucination checking through an additional GPT 4o-Mini
call; therefore, the energy consumption is estimated using the same
approach as for \(E_{\mathrm{GPT\text{ }4o\text{ }mini}}\).


\begin{align}
E_{\mathrm{CNZ},\text{ }hc}\,(\mathrm{kWh})
&= T_{hc}\,(\mathrm{h}) \cdot 0.0085\,(\mathrm{kW}) \cdot 1.09 \\
&= T_{hc}\,(\mathrm{h}) \cdot 0.009265 \nonumber \\
\\ \nonumber
E_{\mathrm{NDC},\text{ }hc}\,(\mathrm{kWh})
&= T_{hc}\,(\mathrm{h}) \cdot 0.343125\,(\mathrm{kW}) \cdot 1.12 \\
&= T_{hc}\,(\mathrm{h}) \cdot 0.3843 \nonumber
\end{align}

\subsection{Evaluation}
To assess the answers, we employ a Human-Expert Evaluation method \cite{belz_comparing_2006}. Evaluation is conducted at the statement level,
treating each statement as the fundamental unit of assessment. When we
identified that the RAG-generated response wasn't relevant or pertinent
to the question asked, we excluded from the response-quality evaluation
for both the RAG system and the pure GPT baseline. Each statement within
a generated response is evaluated in two aspects: Factual Accuracy and
Embellishment Prevalence.

\textbf{Factual Evaluation}: For statements containing verifiable claims
or ground-truth-dependent information, a binary scoring system is
applied. A statement receives a score of 1 (Correct) if and only if it
is entirely consistent with the ground truth; otherwise, it receives a
score of 0. The aggregated Factual Index for each answer is then
calculated as the ratio between the number of correct factual statements
and the total number of factual statements.

\textbf{Embellishment Evaluation:} For each answer we identified the
number of Non-Factual statements, defined as statements that are not
associated to a validated source. The embellishments index for each
answer is calculated as the ratio between the number of non-factual
statements to the total number of statements in the answer.

\section{Results}
\subsection{Energy Consumption}

\begin{figure}[t!]
\centering
\includegraphics[height=3in]{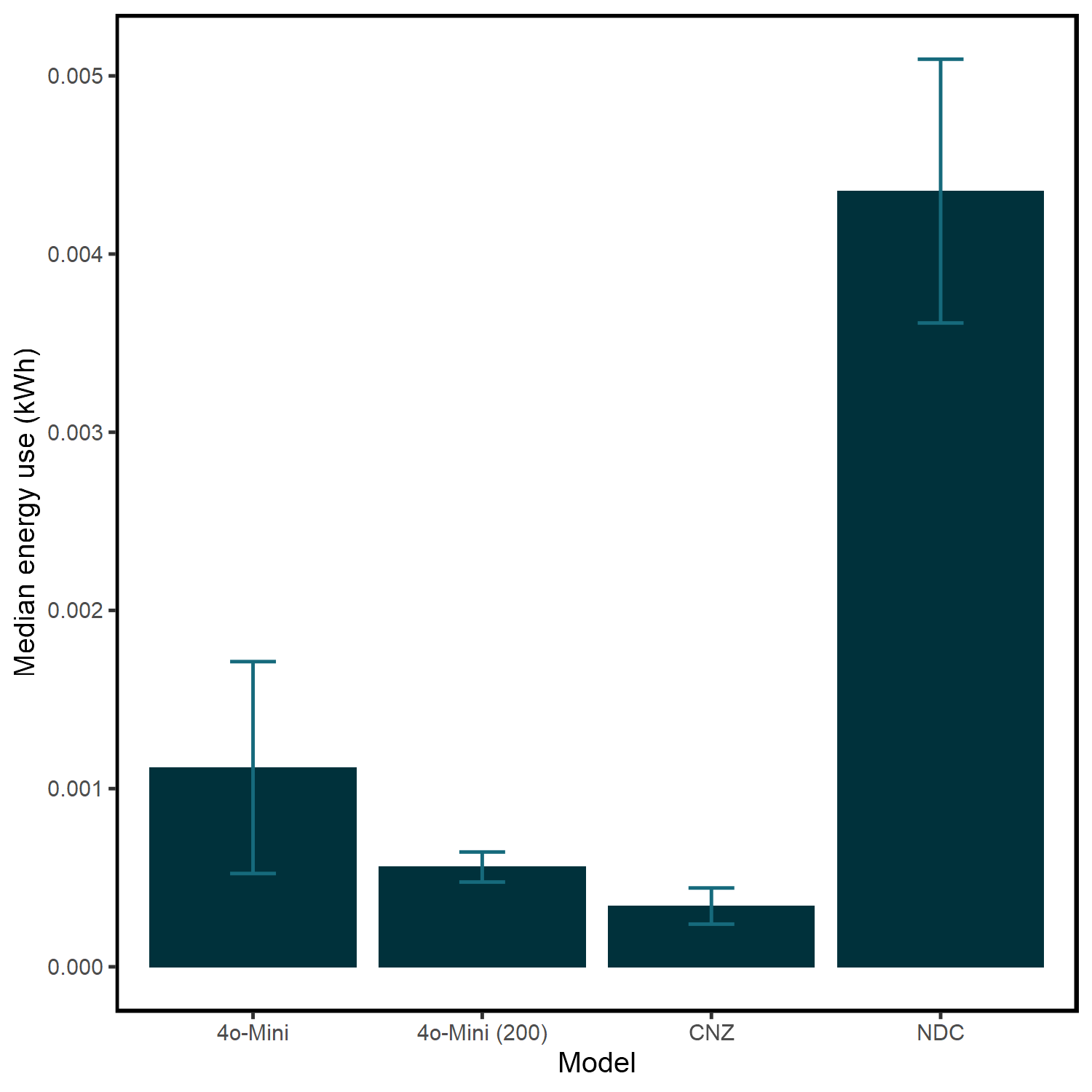}
\caption{Median energy use of models per query.}
\label{fig:med_energy}
\end{figure}

Our results indicate that models vary in terms of their median energy
consumption, with the climate-domain specific chatbots CNZ and
ChatNDC not necessarily consuming less energy compared to GPT 4o-Mini.
Figure~\ref{fig:med_energy} shows the median energy consumption per successful query across
the four model configurations evaluated. Among all systems, ChatNDC had
the highest energy use, consuming \num{4.53e-3} kWh per query. This
consumption was substantially higher than GPT 4o-Mini, which consumed
\num{1.13e-3} kWh, and less energy when constrained to a response of 200
words or less (4o-200: \num{5.0e-4} kWh). CNZ had the lowest mean energy
consumption per query (\num{4.08e-4} kWh) and the smallest variation, with a
mean absolute deviation of \num{1.01e-4}. In contrast, ChatNDC showed both the
highest average consumption and the largest variability, consistent with
a workflow that includes multiple model calls and a more elaborate
post-retrieval verification process compared to CNZ.

\begin{figure}[t!]
\centering
\includegraphics[height=3in]{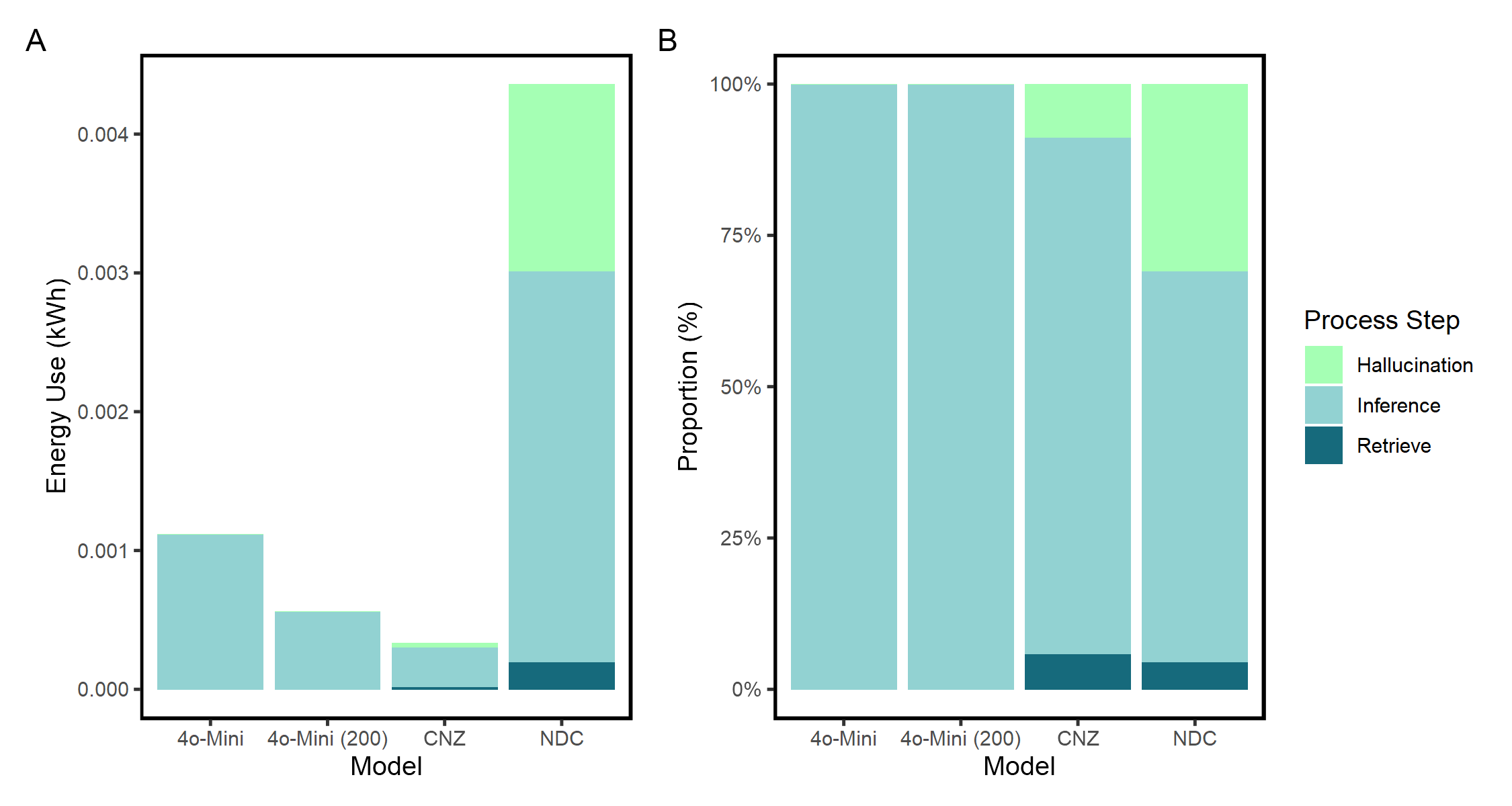}
\caption{Energy use across query steps: (a) Mean energy use of models by each step of the query; (b) Proportion of energy use by step of the query.}
\label{fig:steps}
\end{figure}

Figure \ref{fig:steps} decomposes total query energy into retrieval, inference, and
hallucination-checking components. Across all models, inference accounts
for the largest share of total energy use. For the pure GPT baselines,
energy consumption was almost entirely attributable to inference, since
we were unable to assess estimates of energy consumption for the
retrieval stages. In the case of CNZ, inference also remained a
dominant component, while retrieval and its cosine-similarity-based
hallucination check contributed only a small share of the total energy
use. This pattern is consistent with CNZ's lightweight
verification design, which does not require an additional LLM call like
ChatNDC.

ChatNDC exhibited a markedly different energy profile from CNZ.
Although inference remained the largest single component of total energy
use in both systems, hallucination checking accounted for a
substantially larger share in ChatNDC. Specifically, ChatNDC's
hallucination-checking stage represented 30.9\% of total energy use,
compared with only 8.4\% in CNZ. This disparity arises because ChatNDC
performs statement-level verification through an additional GPT 4o-Mini
call, whereas CNZ relies on a simpler, non-agentic
cosine-similarity method. As a result, ChatNDC's hallucination consumed
approximately 45 times more energy than CNZ's lighter-weight
algorithm. Retrieval, by contrast, remained the smallest energy
component in both RAG systems, and was similar in proportional terms
across CNZ and ChatNDC. These results show that the main difference in
energy use between the two RAG pipelines lies not in retrieval, but in
the much more energy-intensive validation stage introduced by LLM-based
hallucination checking.

The retrieval and hallucination steps observed in CNZ and NDC queries
are part of the additional RAG processes in their architectures.
Although GPT models can search the internet when a topic is outside
their training data, the API does not expose explicit retrieval or
hallucination steps that can be recorded.

\begin{figure}[!b]
\centering
\includegraphics[height=3in,clip,trim=10pt 6pt 6pt 6pt]{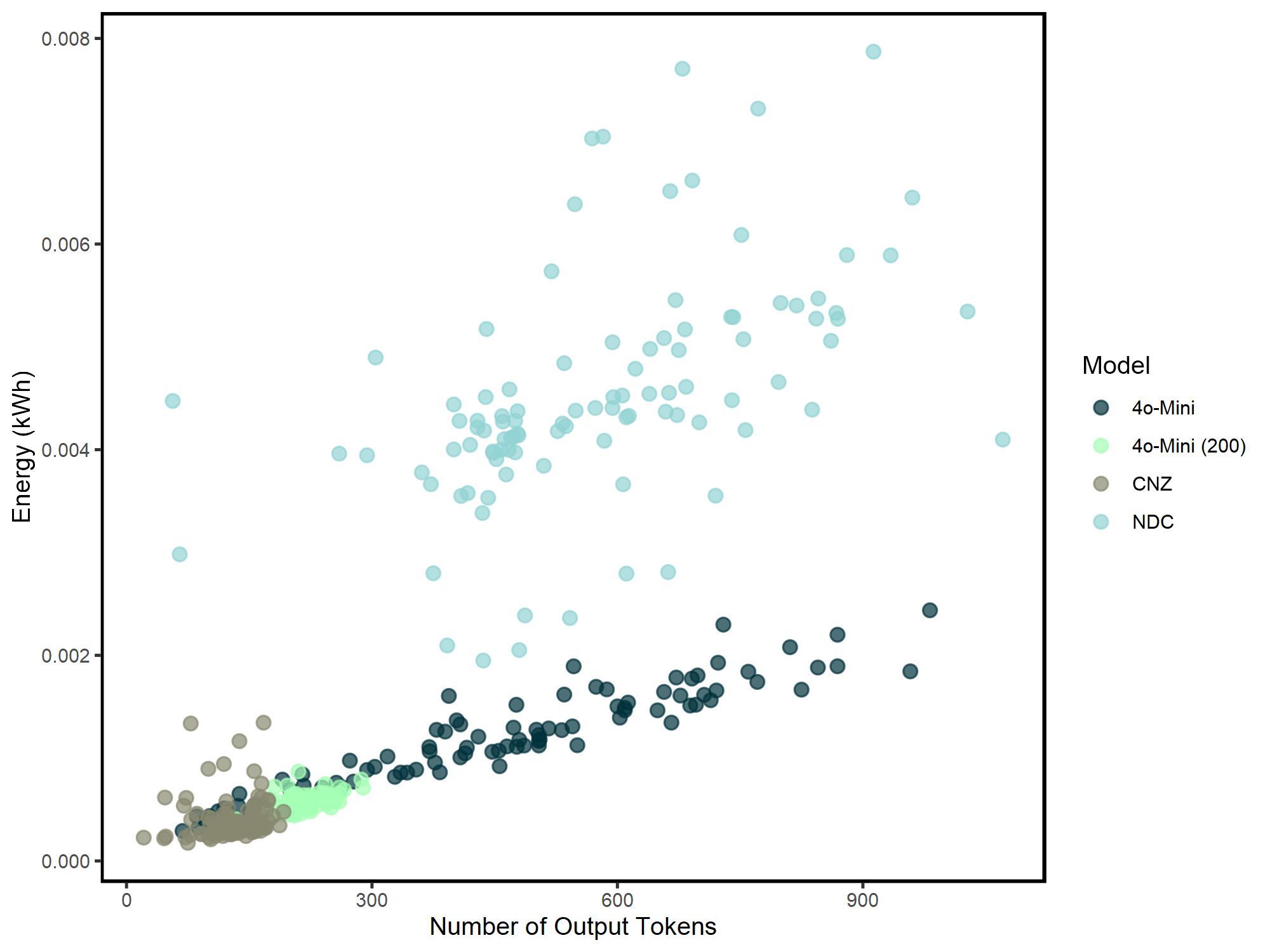}
\caption{Output token vs Energy Usage by model.}
\label{fig:energy_token}
\end{figure}

Exploring the relation between output token size and Energy consumption (Figure \ref{fig:energy_token}),
we observe a particularly strong relation between both variables for the
answers generated by CNZ, GPT 4o-Mini and 4o-Mini (200). To the contrary
the results of ChatNDC shows a weaker association between output token
size and energy consumption. These results highlight not only the direct
relation between these two variables, but also the potential effect of
truncating the answers for increased energy efficiency. Finally, it also
highlights that depending on the system architecture that relation can
be modified such as the case of ChatNDC where output length is not
strongly corrected to energy, due to the additional steps required by
the model and its configuration.

\begin{figure}[t]
\centering
    \centering
    \includegraphics[width=\linewidth,trim=5pt 5pt 5pt 5pt,clip]{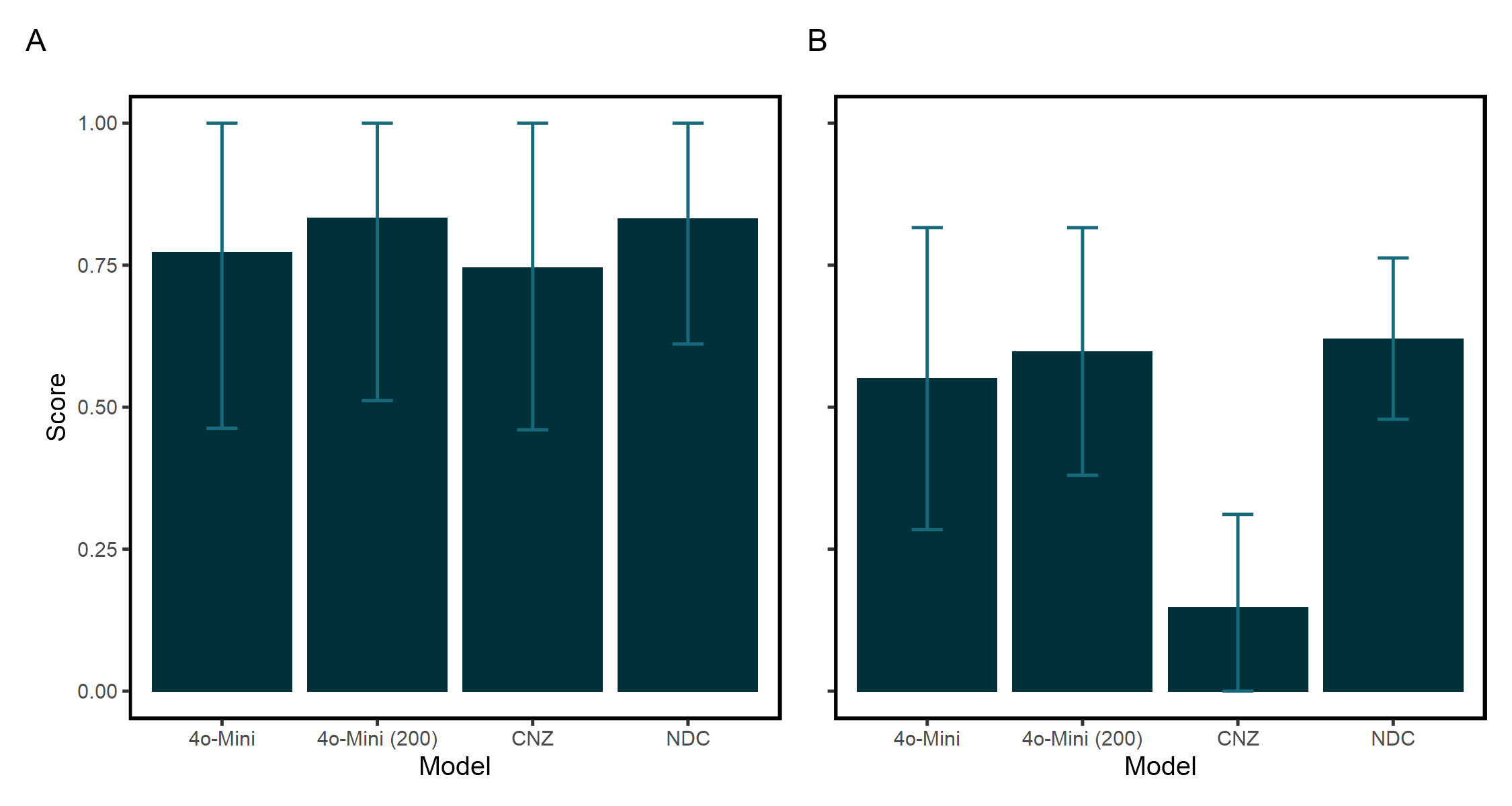}
\caption{a) Mean factual score by model. b) Mean embellishment score by model.}
\label{fig:average_scores}
\end{figure}

Figure \ref{fig:average_scores} shows that mean factual scores are broadly similar across
models, with only modest differences relative to the variation across
questions. By contrast, embellishment differs more clearly by
system. CNZ has the lowest mean embellishment score of CNZ (M = 
0.15; SD = 0.16), which is notably significantly lower than the three
other models (GPT 4o-Mini (M = 0.55, SD = 0.27); GPT 4o-Mini (200) (M = 0.59,
SD = 0.22), and ChatNDC (M = 0.62, SD = 0.14)). This result suggests that CNZ
produces more concise and tightly grounded responses, while the other
systems tend to generate more elaborate outputs without clear gains in
factual performance.

\begin{figure}[t]
\centering
\includegraphics[height=3in]{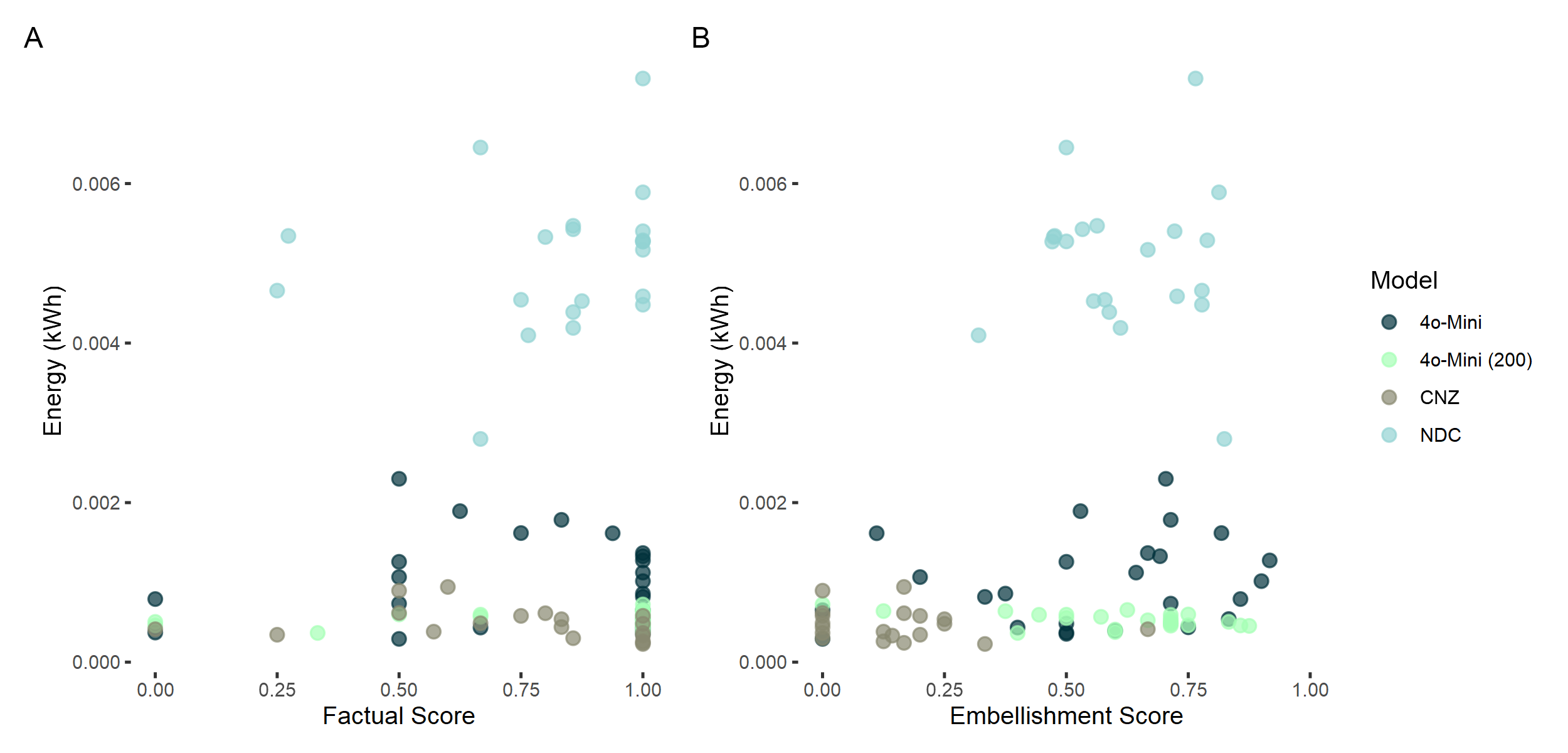}
\caption{a) Energy use vs. factual score of questions, by model. b) Energy use vs. embellishment score of questions, by model.}
\label{fig:factual_embellishment}
\end{figure}

Figure \ref{fig:factual_embellishment} reinforces this interpretation. There is no clear evidence that
higher energy use is associated with higher factual scores. Instead,
systems with substantially greater energy consumption, especially
ChatNDC, achieve factual performance that is broadly similar to
lower-energy alternatives whereas CNZ combines low energy usage with low
embellishment.

Figure \ref{fig:factual_embellishment} explores the association between embellishment, factual
accuracy and output length, as measured by token size. Shorter outputs
do not necessarily guarantee reduced embellishment, as observed in GPT
4o-Mini (200)'s relatively high embellishment score (score = 0.59). CNZ,
however, which is prompted to keep responses under 150 words \cite{hsu_evaluating_2024}, had a much lower embellishment score (0.15). Longer
responses, such as those generated by GPT 4o-Mini and ChatNDC, were more
often associated with higher embellishment, particularly ChatNDC. In
ChatNDC's case, this result appears to stem partly from its system
design, which requires responses between 400-600 words in a
research-paper style. For instance, questions such as ``How many NDCs has Brazil submitted'' that could be answered directly in a single
sentence were often followed by additional and sometimes embellished
content to satisfy the target length. This result suggests that although
RAG-based architectures can help reduce non-factual statements, those
gains may be offset when models are required to produce unnecessarily
long answers.

\begin{figure}[t]
\centering
\includegraphics[width=6.5in,height=2.59722in]{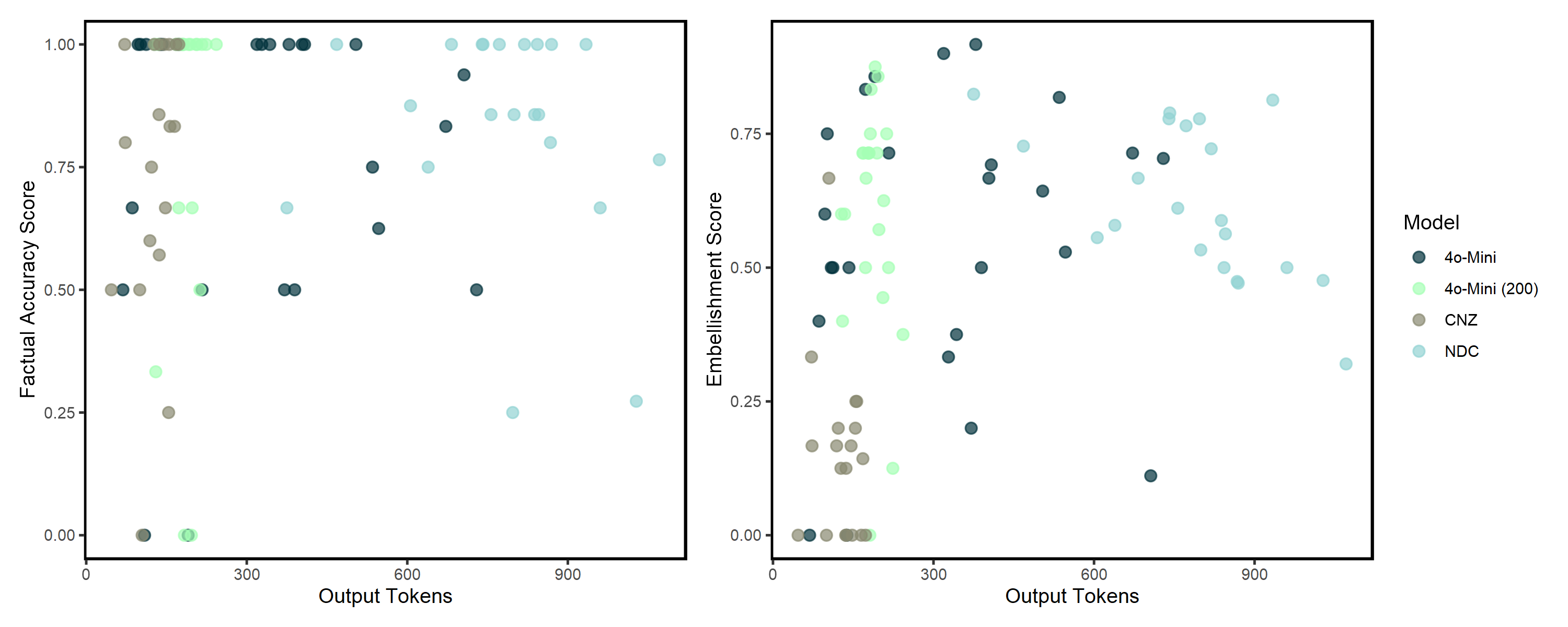}
\caption{a) Output token size vs. factual score of questions, by model. b) Output token size vs. embellishment score of questions, by model.}
\label{fig:output_token}
\end{figure}

\section{Discussion}
As interest in the environmental sustainability of LLMs has grown,
increasing attention has been paid to the energy costs associated with
their real-world deployment. Simultaneously, RAG systems have been
promoted as a promising approach for improving response accuracy,
grounding model outputs in external sources, and reducing hallucinations
in domain-specific applications \cite{hsu_evaluating_2024}. Yet an important
question this study attempts to address is how the energy consumption of
domain-specific RAG systems compare with the direct use of generic LLMs.

Our findings show that while RAG can improve factual grounding, it
introduces additional computational steps, such as retrieval, routing,
and post-generation verification that may increase inference-time energy
use. While domain-specific grounding may allow smaller or more
constrained models to perform competitively without relying on more
expensive generic model outputs, understanding this trade-off is
essential for evaluating not only the quality benefits of RAG, but also
its implications for the sustainable design of LLM-based applications.

A central implication of our findings is that inference-time energy use
is shaped not only by the underlying model, but also by the amount of
text a system is designed to generate. Prior work shows that inference
energy rises with both input and output length, with output length often
exerting a particularly strong influence because each newly generated
token incurs additional decoding steps. Controlled experiments by Poddar
et al.~\cite{poddar_towards_nodate} show near-linear energy growth with output length and also
find that, when input and output lengths are held constant, task
complexity itself has only a small effect on energy use. Luccioni et al.~\cite{luccioni_power_2024} likewise emphasize that generative tasks are substantially more
energy-intensive than discriminative ones, in part because generation
requires repeated token production rather than a single prediction. In
this context, our results suggest that output-length decisions are not
only an energy issue but also a quality issue: systems that encourage
longer or more elaborated answers may create more embellishment,
reflected in the higher embellishment scores. We do not interpret this
as evidence that long answers are inherently less factual, but it does
suggest that expanding output length can increase both energy demand and
the space for non-essential or weakly grounded content.

Our comparison between CNZ and ChatNDC further shows that system
design choices can have large consequences for energy consumption even
when systems rely on the same base generation model. Both products are
RAG systems, yet their energy profiles diverge sharply because of
differences in workflow structure. In our experiments, the main
distinction was not retrieval itself, which remained a relatively small
share of total energy in both systems, but the additional inference
steps introduced by ChatNDC's more agentic verification process. This
result is consistent with the broader agentic RAG literature, which
describes agentic pipelines as adding planning, reflection, iterative
refinement, and tool use on top of standard retrieval and generation \cite{singh_agentic_2025}. Such steps may improve flexibility and robustness,
but they also add inference-time compute through additional model calls.
Our results reinforce a key design point: from an energy perspective,
the major cost of agentic RAG may lie less in retrieval than in the
accumulation of extra post-retrieval reasoning and validation steps.

These findings point toward a more selective design strategy for future
RAG systems. Rather than applying the same high-cost workflow to every
query, developers may achieve better energy efficiency by routing
questions according to their complexity and information requirements.
Recent work on LLM routing \cite{ong_routellm_2025} and cascades \cite{chen_frugalgpt_2023} shows that simpler queries can often be handled by smaller or
cheaper models, while more complex questions are reserved for stronger
models, thereby reducing cost without large quality losses. Relatedly,
sparse mixture-of-experts architectures offer another design path by
activating only a subset of model parameters for each token, which can
improve capability while limiting active compute relative to dense
models \cite{jacobs_adaptive_1991, shazeer_outrageously_2017}. For domain-specific
RAG systems, these suggest a practical next step: simple factual
retrieval questions could be handled by a lightweight pipeline, whereas
open-ended, comparative, or reasoning-heavy questions could trigger a
larger model or a more agentic verification stage. When developers
design domain-specific RAG systems, energy should become an explicit
consideration alongside factual accuracy, latency, and usability.

\section{Limitations}
This study is not without its limitations. First, our energy estimates
rely on end-to-end response time rather than direct server-side response
time from the model provider. This challenge is a common constraint in
API-based LLM research, since proprietary services provide token counts
and some response lifecycle metadata but generally do not provide
detailed server-side information needed for more transparent physical
measurement. Recent work on API-based energy estimation highlights this
problem, noting that API-served LLMs largely remain ``black boxes''
whose energy costs depend heavily on what providers choose to disclose \cite{krupp_this_2026}. Therefore, our estimates should be interpreted as
application-level approximations rather than direct measurements of
energy consumption.

Second, our experiments are limited to single question-response
interactions. While this design improves experimental control, it
doesn't fully reflect how LLM applications are used in practice, where
users may input follow-up prompts in multiple rounds of inference. In
addition, LLM applications tend to incur live search during user
conversations, triggering additional energy usage that is not reflected
in this research when using generic LLM API calls. Future work should
extend analysis from per-query energy estimation to per-session energy
to measure how repeated prompting, conversational memory, and iterative
refinement alter both total consumption and the energy-quality trade-off
in generated responses. This extension is important because emerging
research on reasoning and test-time compute suggests that additional
inference-time computation can substantially increase energy demand as
more tokens or steps are introduced.

Our study also prioritizes methodological refinement over broad
benchmarking across multiple models. We intentionally held the core
model family relatively constant (GPT) in order to isolate the effects
of workflow design, but this choice limits the generalizability of our
absolute energy estimates across providers, model architectures, and
hardware stacks. Prior benchmarking studies have shown substantial
variation in inference impacts across commercial APIs, and public
benchmarking resources such as the ML.ENERGY Leaderboard now cover
dozens of model-task-hardware configurations, illustrating how strongly
efficiency can differ across models and deployment setups \cite{jegham_how_2025}. Future work could therefore expand the empirical scope of
this framework by testing a larger set of LLMs, broader question
datasets, and more access locations in order to improve external
validity and support cross-model comparisons.

Last, both CNZ and ChatNDC rely on static, linear RAG workflows
that lack a mechanism to calibrate computational budgets relative to
query complexity, resulting in energy redundancy for trivial tasks and
insufficiency for sophisticated reasoning. To address these limitations,
future research should pivot toward ``Agentic Workflow'' \cite{yao_react_2023, ye_proagent_2023, wang_internet_2026} frameworks that utilize autonomous
decision chains for elastic reasoning. Such frameworks can dynamically
modulate search breadth and reflection depth, transitioning energy
consumption from rigid, fixed loads to adaptive, flexible loads. Also,
integrating a Sparse Mixture-of-Experts (MoE) architecture offers a
synergistic pathway. These system designs aim to enhance the flexibility
of previous RAGs, balancing energy awareness with high-performance
computing.

\section{Conclusion}
This study examined the inference-time energy consumption of climate
domain-specific LLM and RAG-based chatbot systems by moving beyond
model-level estimates to workflow-level evaluation under real user
queries. Our results show that energy use is shaped not only by the
underlying model, but also by system design choices, including retrieval
structure, output length, and hallucination-checking method. In
particular, more agentic RAG workflows can substantially increase energy
consumption without necessarily producing proportional gains in response
quality in the evaluation setting examined here. Our findings also
suggest that domain-specific RAG systems are not inherently more or less
energy-intensive than generic LLM use, but their energy profile depends
on how the workflow is constructed. By jointly evaluating energy
consumption and answer quality, this study contributes a more
application-relevant framework for assessing the sustainability
trade-offs of LLM-based chatbot design. Future research should extend
this approach to multi-prompt and follow-up interactions, a wider range
of models and locations, and more dynamic agentic workflows.

\section*{Acknowledgements}
This work was supported by grants from the IKEA Foundation and William and Flora Hewlett Foundation to A.Hsu. The authors acknowledge the contributions of Mariam Maher, Arboretica, in the algorithmic design of ChatNetZero. The development of ChatNDC was supported by an award from the Climateworks Foundation, and the University of North Carolina at Chapel Hill supported the development of ChatNetZero through a seed grant to A. Hsu.

\bibliographystyle{unsrt}
\bibliography{r1, r2}

@article{vaghefi_chatipcc_2023,
	title = {{ChatIPCC}: {Grounding} {Conversational} {AI} in {Climate} {Science}},
	issn = {1556-5068},
	shorttitle = {{ChatIPCC}},
	url = {https://www.ssrn.com/abstract=4414628},
	doi = {10.2139/ssrn.4414628},
	language = {en},
	urldate = {2023-08-14},
	journal = {SSRN Electronic Journal},
	author = {Vaghefi, Saeid and Wang, Qian and Muccione, Veruska and Ni, Jingwei and Kraus, Mathias and Bingler, Julia and Schimanski, Tobias and Colesanti Senni, Chiara and Webersinke, Nicolas and Huggel, Christian and Leippold, Markus},
	year = {2023},
}

@article{samborska_scaling_2025,
	title = {Scaling up: how increasing inputs has made artificial intelligence more capable},
	shorttitle = {Scaling up},
	url = {https://ourworldindata.org/scaling-up-ai},
	language = {en},
	urldate = {2026-03-05},
	journal = {Our World in Data},
	author = {Samborska, Veronika},
	month = jan,
	year = {2025},
}

@misc{vrettos_accurate_nodate,
	title = {Accurate and {Energy} {Efficient}: {Local} {Retrieval}- {Augmented} {Generation} {Models} {Outperform} {Commercial} {Large} {Language} {Models} in {Medical} {Tasks}},
	language = {en},
	author = {Vrettos, Konstantinos and Klontzas, Michail E},
}

@article{shahbazi_using_2026,
	title = {Using {Large} {Language} {Models} to {Detect} and {Debunk} {Climate} {Change} {Misinformation}},
	volume = {10},
	copyright = {http://creativecommons.org/licenses/by/3.0/},
	issn = {2504-2289},
	url = {https://www.mdpi.com/2504-2289/10/1/34},
	doi = {10.3390/bdcc10010034},
	language = {en},
	number = {1},
	urldate = {2026-03-05},
	journal = {Big Data and Cognitive Computing},
	author = {Shahbazi, Zeinab and Behnamian, Sara},
	month = jan,
	year = {2026},
	note = {Publisher: Multidisciplinary Digital Publishing Institute},
}

@misc{meredith_thirsty_2023,
	title = {A ‘thirsty’ generative {AI} boom poses a growing problem for {Big} {Tech}},
	url = {https://www.cnbc.com/2023/12/06/water-why-a-thirsty-generative-ai-boom-poses-a-problem-for-big-tech.html},
	language = {en},
	urldate = {2026-03-05},
	journal = {CNBC},
	author = {Meredith, Sam},
	month = dec,
	year = {2023},
	note = {Section: COP28},
}

@article{copley_data_2025,
	chapter = {Climate},
	title = {Data centers are booming. {But} there are big energy and environmental risks},
	url = {https://www.npr.org/2025/10/14/nx-s1-5565147/google-ai-data-centers-growth-environment-electricity},
	language = {en},
	urldate = {2026-03-05},
	journal = {NPR},
	author = {Copley, Michael},
	month = oct,
	year = {2025},
}

@misc{iea_energy_2025,
	title = {Energy demand from {AI} – {Energy} and {AI} – {Analysis}},
	url = {https://www.iea.org/reports/energy-and-ai/energy-demand-from-ai},
	language = {en-GB},
	urldate = {2026-03-05},
	journal = {IEA},
	author = {IEA},
	year = {2025},
}

@article{chen_how_2025,
	title = {How much energy will {AI} really consume? {The} good, the bad and the unknown},
	volume = {639},
	copyright = {2025 Springer Nature Limited},
	issn = {1476-4687},
	shorttitle = {How much energy will {AI} really consume?},
	url = {https://www.nature.com/articles/d41586-025-00616-z},
	doi = {10.1038/d41586-025-00616-z},
	language = {en},
	number = {8053},
	urldate = {2026-03-05},
	journal = {Nature},
	author = {Chen, Sophia},
	month = mar,
	year = {2025},
	note = {Bandiera\_abtest: a
Cg\_type: News Feature
Publisher: Nature Publishing Group
Subject\_term: Machine learning, Computer science, Sustainability, Energy},
	pages = {22--24},
}

@misc{li_making_2025,
	title = {Making {AI} {Less} "{Thirsty}": {Uncovering} and {Addressing} the {Secret} {Water} {Footprint} of {AI} {Models}},
	shorttitle = {Making {AI} {Less} "{Thirsty}"},
	url = {http://arxiv.org/abs/2304.03271},
	doi = {10.48550/arXiv.2304.03271},
	urldate = {2026-03-05},
	publisher = {arXiv},
	author = {Li, Pengfei and Yang, Jianyi and Islam, Mohammad A. and Ren, Shaolei},
	month = mar,
	year = {2025},
	note = {arXiv:2304.03271 [cs]},
}

@misc{jegham_how_2025,
	title = {How {Hungry} is {AI}? {Benchmarking} {Energy}, {Water}, and {Carbon} {Footprint} of {LLM} {Inference}},
	shorttitle = {How {Hungry} is {AI}?},
	url = {http://arxiv.org/abs/2505.09598},
	doi = {10.48550/arXiv.2505.09598},
	urldate = {2026-03-16},
	publisher = {arXiv},
	author = {Jegham, Nidhal and Abdelatti, Marwan and Koh, Chan Young and Elmoubarki, Lassad and Hendawi, Abdeltawab},
	month = nov,
	year = {2025},
	note = {arXiv:2505.09598 [cs]},
}

@misc{lacoste_quantifying_2019,
	title = {Quantifying the {Carbon} {Emissions} of {Machine} {Learning}},
	url = {http://arxiv.org/abs/1910.09700},
	doi = {10.48550/arXiv.1910.09700},
	urldate = {2026-03-16},
	publisher = {arXiv},
	author = {Lacoste, Alexandre and Luccioni, Alexandra and Schmidt, Victor and Dandres, Thomas},
	month = nov,
	year = {2019},
	note = {arXiv:1910.09700 [cs]},
}

@inproceedings{everman_evaluating_2023,
	title = {Evaluating the carbon impact of large language models at the inference stage},
	url = {https://ieeexplore.ieee.org/abstract/document/10253886/},
	urldate = {2026-03-17},
	booktitle = {2023 {IEEE} international performance, computing, and communications conference ({IPCCC})},
	publisher = {IEEE},
	author = {Everman, Brad and Villwock, Trevor and Chen, Dayuan and Soto, Noe and Zhang, Oliver and Zong, Ziliang},
	year = {2023},
	pages = {150--157},
}

@article{dauner_energy_2025,
	title = {Energy costs of communicating with {AI}},
	volume = {10},
	url = {https://www.frontiersin.org/journals/communication/articles/10.3389/fcomm.2025.1572947/full},
	urldate = {2026-03-17},
	journal = {Frontiers in Communication},
	author = {Dauner, Maximilian and Socher, Gudrun},
	year = {2025},
	note = {Publisher: Frontiers Media SA},
	pages = {1572947},
}

@article{nguyen_towards_2024,
	title = {Towards {Sustainable} {Large} {Language} {Model} {Serving}},
	volume = {4},
	issn = {2770-5331, 2770-5331},
	url = {https://dl.acm.org/doi/10.1145/3727200.3727220},
	doi = {10.1145/3727200.3727220},
	language = {en},
	number = {5},
	urldate = {2026-03-17},
	journal = {ACM SIGEnergy Energy Informatics Review},
	author = {Nguyen, Sophia and Zhou, Beihao and Ding, Yi and Liu, Sihang},
	month = dec,
	year = {2024},
	pages = {134--140},
}

@inproceedings{sikand_breaking_2025,
	title = {Breaking the {ICE}: {Exploring} promises and challenges of benchmarks for {Inference} {Carbon} \& {Energy} estimation for {LLMs}},
	shorttitle = {Breaking the {ICE}},
	url = {http://arxiv.org/abs/2506.08727},
	doi = {10.1109/GREENS66463.2025.00013},
	urldate = {2026-03-17},
	booktitle = {2025 {IEEE}/{ACM} 9th {International} {Workshop} on {Green} and {Sustainable} {Software} ({GREENS})},
	author = {Sikand, Samarth and Mehra, Rohit and Pathania, Priyavanshi and Bamby, Nikhil and Sharma, Vibhu Saujanya and Kaulgud, Vikrant and Podder, Sanjay and Burden, Adam P.},
	month = apr,
	year = {2025},
	note = {arXiv:2506.08727 [cs]},
	pages = {55--59},
}

@misc{husom_price_2026,
	title = {The {Price} of {Prompting}: {Profiling} {Energy} {Use} in {Large} {Language} {Models} {Inference}},
	shorttitle = {The {Price} of {Prompting}},
	url = {http://arxiv.org/abs/2407.16893},
	doi = {10.48550/arXiv.2407.16893},
	urldate = {2026-03-17},
	publisher = {arXiv},
	author = {Husom, Erik Johannes and Goknil, Arda and Shar, Lwin Khin and Sen, Sagar},
	month = mar,
	year = {2026},
	note = {arXiv:2407.16893 [cs]},
}

@article{odonnell_we_2025,
	title = {We did the math on {AI}’s energy footprint. {Here}’s the story you haven’t heard.},
	url = {https://www.technologyreview.com/2025/05/20/1116327/ai-energy-usage-climate-footprint-big-tech/},
	language = {en},
	urldate = {2026-03-18},
	journal = {MIT Technology Review},
	author = {O'Donnell, James and Crownhart, Casey},
	month = may,
	year = {2025},
}

@inproceedings{luccioni_power_2024,
	title = {Power {Hungry} {Processing}: {Watts} {Driving} the {Cost} of {AI} {Deployment}?},
	shorttitle = {Power {Hungry} {Processing}},
	url = {http://arxiv.org/abs/2311.16863},
	doi = {10.1145/3630106.3658542},
	urldate = {2026-03-18},
	booktitle = {The 2024 {ACM} {Conference} on {Fairness} {Accountability} and {Transparency}},
	author = {Luccioni, Alexandra Sasha and Jernite, Yacine and Strubell, Emma},
	month = jun,
	year = {2024},
	note = {arXiv:2311.16863 [cs]},
	pages = {85--99},
}

@article{ormell_blooms_1974,
	title = {Bloom's {Taxonomy} and the {Objectives} of {Education}},
	volume = {17},
	issn = {0013-1881, 1469-5847},
	url = {https://www.tandfonline.com/doi/full/10.1080/0013188740170101},
	doi = {10.1080/0013188740170101},
	language = {en},
	number = {1},
	urldate = {2026-03-18},
	journal = {Educational Research},
	author = {Ormell, C. P.},
	month = nov,
	year = {1974},
	pages = {3--18},
}

@misc{intel_intel_nodate,
	title = {Intel {Xeon} {Processor}},
	url = {https://www.intel.com/content/www/us/en/products/details/processors/xeon.html},
	journal = {Intel},
	author = {Intel},
}

@misc{google_power_nodate,
	title = {Power usage effectiveness – {Google} {Data} {Centers}},
	url = {https://datacenters.google/efficiency},
	urldate = {2026-03-19},
	journal = {Google Data Centers},
	author = {Google},
}

@misc{faiz_llmcarbon_2024,
	title = {{LLMCarbon}: {Modeling} the end-to-end {Carbon} {Footprint} of {Large} {Language} {Models}},
	shorttitle = {{LLMCarbon}},
	url = {http://arxiv.org/abs/2309.14393},
	doi = {10.48550/arXiv.2309.14393},
	urldate = {2026-03-19},
	publisher = {arXiv},
	author = {Faiz, Ahmad and Kaneda, Sotaro and Wang, Ruhan and Osi, Rita and Sharma, Prateek and Chen, Fan and Jiang, Lei},
	month = jan,
	year = {2024},
	note = {arXiv:2309.14393 [cs]},
}

@article{fu_llmco2_2025,
	title = {{LLMCO}$_{\textrm{2}}$ : {Advancing} {Accurate} {Carbon} {Footprint} {Prediction} for {LLM} {Inferences}},
	volume = {5},
	issn = {2770-5331, 2770-5331},
	shorttitle = {{LLMCO}$_{\textrm{2}}$},
	url = {https://dl.acm.org/doi/10.1145/3757892.3757901},
	doi = {10.1145/3757892.3757901},
	language = {en},
	number = {2},
	urldate = {2026-03-19},
	journal = {ACM SIGEnergy Energy Informatics Review},
	author = {Fu, Zhenxiao and Chen, Fan and Zhou, Shan and Li, Haitong and Jiang, Lei},
	month = jul,
	year = {2025},
	pages = {63--68},
}

@inproceedings{poddar_towards_nodate,
    title = {Towards {Sustainable} {NLP}: {Insights} from {Benchmarking} {Inference} {Energy} in {Large} {Language} {Models}},
    author = {Poddar, Soham and Koley, Paramita and Misra, Janardan and Ganguly, Niloy and Ghosh, Saptarshi},
    booktitle = {Proceedings of the 2025 Conference of the Nations of the Americas Chapter of the Association for Computational Linguistics: Human Language Technologies (Volume 1: Long Papers)},
    pages = {12688--12704},
    year = {2025},
    address = {Albuquerque, New Mexico},
    publisher = {Association for Computational Linguistics},
}

@misc{singh_agentic_2025,
	title = {Agentic {Retrieval}-{Augmented} {Generation}: {A} {Survey} on {Agentic} {RAG}},
	shorttitle = {Agentic {Retrieval}-{Augmented} {Generation}},
	url = {http://arxiv.org/abs/2501.09136},
	doi = {10.48550/arXiv.2501.09136},
	urldate = {2026-03-19},
	publisher = {arXiv},
	author = {Singh, Aditi and Ehtesham, Abul and Kumar, Saket and Khoei, Tala Talaei},
	month = feb,
	year = {2025},
	note = {arXiv:2501.09136 [cs]},
}

@misc{ong_routellm_2025,
	title = {{RouteLLM}: {Learning} to {Route} {LLMs} with {Preference} {Data}},
	shorttitle = {{RouteLLM}},
	url = {http://arxiv.org/abs/2406.18665},
	doi = {10.48550/arXiv.2406.18665},
	urldate = {2026-03-19},
	publisher = {arXiv},
	author = {Ong, Isaac and Almahairi, Amjad and Wu, Vincent and Chiang, Wei-Lin and Wu, Tianhao and Gonzalez, Joseph E. and Kadous, M. Waleed and Stoica, Ion},
	month = feb,
	year = {2025},
	note = {arXiv:2406.18665 [cs]},
}

@misc{chen_frugalgpt_2023,
	title = {{FrugalGPT}: {How} to {Use} {Large} {Language} {Models} {While} {Reducing} {Cost} and {Improving} {Performance}},
	shorttitle = {{FrugalGPT}},
	url = {http://arxiv.org/abs/2305.05176},
	doi = {10.48550/arXiv.2305.05176},
	urldate = {2026-03-19},
	publisher = {arXiv},
	author = {Chen, Lingjiao and Zaharia, Matei and Zou, James},
	month = may,
	year = {2023},
	note = {arXiv:2305.05176 [cs]},
}

@misc{krupp_this_2026,
	title = {This {Is} {Taking} {Too} {Long} -- {Investigating} {Time} as a {Proxy} for {Energy} {Consumption} of {LLMs}},
	url = {http://arxiv.org/abs/2603.15699},
	doi = {10.48550/arXiv.2603.15699},
	urldate = {2026-03-19},
	publisher = {arXiv},
	author = {Krupp, Lars and Geißler, Daniel and Calatrava-Nicolas, Francisco M. and Banwari, Vishal and Lukowicz, Paul and Karolus, Jakob},
	month = mar,
	year = {2026},
	note = {arXiv:2603.15699 [cs]
version: 1},
}

@misc{ariyanti_sri_trade-off_nodate,
	title = {Trade-{Off} {Between} {Energy} {Consumption} and {Three} {Configuration} {Parameters} in {Artificial} {Intelligence} ({AI}) {Training}: {Lessons} for {Environmental} {Policy}},
	url = {https://www.mdpi.com/2071-1050/17/12/5359},
	urldate = {2026-03-23},
	author = {{Ariyanti Sri} and {Suryanegara Muhammad}},
}

@inproceedings{belz_comparing_2006,
	address = {Trento, Italy},
	title = {Comparing {Automatic} and {Human} {Evaluation} of {NLG} {Systems}},
	url = {https://aclanthology.org/E06-1040/},
	urldate = {2026-03-23},
	booktitle = {11th {Conference} of the {European} {Chapter} of the {Association} for {Computational} {Linguistics}},
	publisher = {Association for Computational Linguistics},
	author = {Belz, Anja and Reiter, Ehud},
	editor = {McCarthy, Diana and Wintner, Shuly},
	month = apr,
	year = {2006},
	pages = {313--320},
}

@misc{cottier_rising_2025,
	title = {The rising costs of training frontier {AI} models},
	url = {http://arxiv.org/abs/2405.21015},
	doi = {10.48550/arXiv.2405.21015},
	urldate = {2026-03-23},
	publisher = {arXiv},
	author = {Cottier, Ben and Rahman, Robi and Fattorini, Loredana and Maslej, Nestor and Besiroglu, Tamay and Owen, David},
	month = feb,
	year = {2025},
	note = {arXiv:2405.21015 [cs]},
}

@article{guardia_assessing_2024,
	title = {Assessing the {Energy} {Impact} and {Carbon} {Footprint} of {AI} {Model} {Training}: {A} {Case} {Study} {Using} {GPU} {Servers}.},
	language = {en},
	author = {Guardia, Giovanni Cocca and Vigneau, Gabriel Hermosilla and Espín-Sarzosa, Danny and Feris, Bárbara Dumas},
    journal = {Resources, Conservation and Recycling},
	year = {2024},
}

@article{jacobs_adaptive_1991,
	title = {Adaptive {Mixtures} of {Local} {Experts}},
	volume = {3},
	issn = {0899-7667},
	url = {https://ieeexplore.ieee.org/abstract/document/6797059},
	doi = {10.1162/neco.1991.3.1.79},
	number = {1},
	urldate = {2026-03-23},
	journal = {Neural Computation},
	author = {Jacobs, Robert A. and Jordan, Michael I. and Nowlan, Steven J. and Hinton, Geoffrey E.},
	month = mar,
	year = {1991},
	pages = {79--87},
}

@article{kaack_aligning_2022,
	title = {Aligning artificial intelligence with climate change mitigation},
	volume = {12},
	issn = {1758-678X, 1758-6798},
	url = {https://www.nature.com/articles/s41558-022-01377-7},
	doi = {10.1038/s41558-022-01377-7},
	language = {en},
	number = {6},
	urldate = {2026-03-23},
	journal = {Nature Climate Change},
	author = {Kaack, Lynn H. and Donti, Priya L. and Strubell, Emma and Kamiya, George and Creutzig, Felix and Rolnick, David},
	month = jun,
	year = {2022},
	pages = {518--527},
}

@misc{patterson_carbon_2021,
	title = {Carbon {Emissions} and {Large} {Neural} {Network} {Training}},
	url = {http://arxiv.org/abs/2104.10350},
	doi = {10.48550/arXiv.2104.10350},
	urldate = {2026-03-23},
	publisher = {arXiv},
	author = {Patterson, David and Gonzalez, Joseph and Le, Quoc and Liang, Chen and Munguia, Lluis-Miquel and Rothchild, Daniel and So, David and Texier, Maud and Dean, Jeff},
	month = apr,
	year = {2021},
	note = {arXiv:2104.10350 [cs]},
}

@misc{shazeer_outrageously_2017,
	title = {Outrageously {Large} {Neural} {Networks}: {The} {Sparsely}-{Gated} {Mixture}-of-{Experts} {Layer}},
	shorttitle = {Outrageously {Large} {Neural} {Networks}},
	url = {http://arxiv.org/abs/1701.06538},
	doi = {10.48550/arXiv.1701.06538},
	urldate = {2026-03-23},
	publisher = {arXiv},
	author = {Shazeer, Noam and Mirhoseini, Azalia and Maziarz, Krzysztof and Davis, Andy and Le, Quoc and Hinton, Geoffrey and Dean, Jeff},
	month = jan,
	year = {2017},
	note = {arXiv:1701.06538 [cs]},
}

@article{wang_internet_2026,
	title = {Internet of {Agents}: {Fundamentals}, {Applications}, and {Challenges}},
	volume = {12},
	issn = {2332-7731, 2372-2045},
	shorttitle = {Internet of {Agents}},
	url = {http://arxiv.org/abs/2505.07176},
	doi = {10.1109/TCCN.2025.3623369},
	urldate = {2026-03-23},
	journal = {IEEE Transactions on Cognitive Communications and Networking},
	author = {Wang, Yuntao and Guo, Shaolong and Pan, Yanghe and Su, Zhou and Chen, Fahao and Luan, Tom H. and Li, Peng and Kang, Jiawen and Niyato, Dusit},
	year = {2026},
	note = {arXiv:2505.07176 [cs]},
	pages = {4476--4501},
}

@misc{yao_react_2023,
	title = {{ReAct}: {Synergizing} {Reasoning} and {Acting} in {Language} {Models}},
	shorttitle = {{ReAct}},
	url = {http://arxiv.org/abs/2210.03629},
	doi = {10.48550/arXiv.2210.03629},
	urldate = {2026-03-23},
	publisher = {arXiv},
	author = {Yao, Shunyu and Zhao, Jeffrey and Yu, Dian and Du, Nan and Shafran, Izhak and Narasimhan, Karthik and Cao, Yuan},
	month = mar,
	year = {2023},
	note = {arXiv:2210.03629 [cs]},
}

@misc{ye_proagent_2023,
	title = {{ProAgent}: {From} {Robotic} {Process} {Automation} to {Agentic} {Process} {Automation}},
	shorttitle = {{ProAgent}},
	url = {http://arxiv.org/abs/2311.10751},
	doi = {10.48550/arXiv.2311.10751},
	urldate = {2026-03-23},
	publisher = {arXiv},
	author = {Ye, Yining and Cong, Xin and Tian, Shizuo and Cao, Jiannan and Wang, Hao and Qin, Yujia and Lu, Yaxi and Yu, Heyang and Wang, Huadong and Lin, Yankai and Liu, Zhiyuan and Sun, Maosong},
	month = nov,
	year = {2023},
	note = {arXiv:2311.10751 [cs]},
}

@misc{noauthor_tiktoken_2026,
	title = {Tiktoken},
	copyright = {MIT},
	url = {https://github.com/openai/tiktoken},
	urldate = {2026-03-23},
	publisher = {OpenAI},
	month = mar,
	year = {2026},
	note = {original-date: 2022-12-01T23:22:11Z},
}

@misc{microsoft_measuring_nodate,
	title = {Measuring energy and water efficiency for {Microsoft} {Datacenters}},
	url = {https://datacenters.microsoft.com/sustainability/efficiency/},
	language = {en-US},
	urldate = {2026-03-26},
	journal = {Microsoft Datacenters},
	author = {Microsoft},
}

@misc{elsworth_measuring_2025,
	title = {Measuring the environmental impact of delivering {AI} at {Google} {Scale}},
	url = {http://arxiv.org/abs/2508.15734},
	doi = {10.48550/arXiv.2508.15734},
	urldate = {2026-03-28},
	publisher = {arXiv},
	author = {Elsworth, Cooper and Huang, Keguo and Patterson, David and Schneider, Ian and Sedivy, Robert and Goodman, Savannah and Townsend, Ben and Ranganathan, Parthasarathy and Dean, Jeff and Vahdat, Amin and Gomes, Ben and Manyika, James},
	month = aug,
	year = {2025},
	note = {arXiv:2508.15734 [cs]},
}

@techreport{net_zero_tracker_net_2025,
	title = {Net {Zero} {Stocktake} 2025},
	url = {https://zerotracker.net/analysis/net-zero-stocktake-2025},
	urldate = {2026-03-28},
	institution = {NewClimate Institute, Oxford Net Zero, Energy and Climate Intelligence Unit, and Data-Driven EnviroLab.},
	author = {Net Zero Tracker},
	year = {2025},
}

@inproceedings{hsu_evaluating_2024,
	address = {Bangkok, Thailand},
	title = {Evaluating {ChatNetZero}, an {LLM}-{Chatbot} to {Demystify} {Climate} {Pledges}},
	url = {https://aclanthology.org/2024.climatenlp-1.6},
	doi = {10.18653/v1/2024.climatenlp-1.6},
	language = {en},
	urldate = {2026-03-28},
	booktitle = {Proceedings of the 1st {Workshop} on {Natural} {Language} {Processing} {Meets} {Climate} {Change} ({ClimateNLP} 2024)},
	publisher = {Association for Computational Linguistics},
	author = {Hsu, Angel and Laney, Mason and Zhang, Ji and Manya, Diego and Farczadi, Linda},
	year = {2024},
	pages = {82--92},
}

@misc{manya-gutierrez_replication_2026,
	title = {Replication {Data} for: {Assessing} the {Energy} {Consumption} of {LLMs} and {Domain} {Products}},
	doi = {https://doi.org/10.15139/S3/PMKQGH},
	author = {Manya-Gutierrez, Diego and Bao, Alicia and He, Jiamian and Hsu, Angel and Zhang, Ji},
	year = {2026},
}
\section*{Supplementary material}
The experimental data are publicly available on the Data-Driven EnviroLab's Dataverse page \cite{manya-gutierrez_replication_2026}. 

\end{document}